\documentclass[twoside,journal]{IEEEtran} 
\usepackage{xcolor,colortbl}
\usepackage[normalem]{ulem}

\newcommand{\subparagraph}{} 
\usepackage{titlesec}
\titlespacing{\section}{0pt}{0.4\baselineskip}{0.20\baselineskip}
\titlespacing{\subsection}{0pt}{0.2\baselineskip}{0.2\baselineskip}

\usepackage{amsmath}
\usepackage{graphicx}
\usepackage{float}
\usepackage{stfloats}
\usepackage[caption=false]{subfig}
\usepackage{svg}

\usepackage{makecell}
\usepackage{multirow}
\usepackage{booktabs}

\usepackage{cite}
\usepackage[colorlinks=false,hidelinks]{hyperref}
\hypersetup{
	bookmarks=true,         
	unicode=false,          
	pdftoolbar=true,        
	pdfmenubar=true,        
	pdffitwindow=false,     
	pdftitle={},    
	pdfauthor={},     
	pdfsubject={},   
	pdfcreator={},   
	pdfproducer={},  
	pdfkeywords={}, 
	pdfnewwindow=true,      
	colorlinks=false       
}

\usepackage{amsthm}
\usepackage{algorithm}
\usepackage{algorithmic}

\usepackage{mathptmx}
\usepackage{newtxmath} 

\usepackage[T1]{fontenc}
\usepackage{aecompl}

\begin{document}

\title{
	Hierarchically Coordinated Energy Management for A Regional Multi-microgrid Community
}

\author{
	Chengquan~Ju
}
\maketitle

\markboth{}{}

\begin{abstract}
	This paper proposes a novel hierarchically coordinated energy management system (EMS) for a regional community (e.g., residential area, campus, industrial park, etc.) comprising multiple small-scale microgrids (MGs) (e.g., houses, buildings, etc.). 
	It aims to minimize the total operational cost of the MG community and maximize the individual benefit of each MG simultaneously. 
	At the local level inside each MG, with the detailed modeling of various energy resources including photovoltaics (PVs), energy storages (ESs), electric vehicles (EVs) and dispatchable loads, the individual optimization problem is formulated as a mixed-integer linear program (MILP). Local EMSs makes power dispatch decisions for all the controllable units to minimize the operational cost in individual MGs. 
	At the community level, a novel pairing algorithm is proposed to explicitly find the MG pairings with surplus and deficit. The community-level EMS employs the pairing algorithm to determine specific power exchanges among MGs and minimizes the energy transactions with the upstream grid. The operational cost of each individual MG is further reduced by additional economic benefits procured by the community-level EMS. 
	The proposed method has distinguishing advantages on modeling generality, computational complexity and privacy protection, and its performance is verified by the simulation results. 
\end{abstract}
\vspace{-0.2cm}
\begin{IEEEkeywords}
	Energy management, microgrids, hierarchical optimization, energy transaction. 
\end{IEEEkeywords}

\section*{Nomenclature}
\addcontentsline{toc}{section}{Nomenclature}
\subsection{Indices and Sets}
\begin{IEEEdescription}[\IEEEusemathlabelsep\IEEEsetlabelwidth{$~~~~~~~~~$}]
	\item[$\Delta t$] Time interval.
	\item[$t$] Index of time.
	\item[$i,j$] Indices of MG. 
	\item[$k$] Index of controllable appliances. 
	\item[$n$] Index of special ordered set of type 2 (SOS-2).

	\item[$\boldsymbol {N_G}$] Set of MGs.
	\item[$\boldsymbol {{ES}^i}$] Set of Energy Storages (ESs) in MG $i$.
	\item[$\boldsymbol {{EV}^i}$] Set of Electric Vehicles (EVs) in MG $i$.
	\item[$\boldsymbol {L_1^i}$] Set of type 1 loads in MG $i$.
	\item[$\boldsymbol {L_2^i}$] Set of type 2 loads in MG $i$.
	\item[$\boldsymbol{N_k^i}$] SOS-2 for $k$th ES in MG $i$.
\end{IEEEdescription}

\subsection{Decision Variables}
\begin{IEEEdescription}[\IEEEusemathlabelsep\IEEEsetlabelwidth{$~~~~~~~~~$}]
	\item[$p_{\!M\!,b}^{i,t},p_{\!M\!,s}^{i,t}$] 	Purchasing and selling power of upstream grid.
	\item[$p_{C,b}^{ij,t},p_{C,s}^{ij,t}$] 	Transmitted power between MG $i$ and $j$. 
	\item[$p_{k,b}^{i,t},p_{k,s}^{i,t}$] 	Discharging and charging power of ES (EV).

	\item[$u_M^{i,t}$] 	Binary indicator for power from utility grid. 
	\item[$u_C^{ij,t}$] 	Binary indicator for transmitted power between $i$th and $j$th MG.
	\item[$\delta_k^{i,t}$] 	Binary indicator for power of ES. 
	\item[$\theta_k^{i,t}$] 	Binary indicator for power of EV. 
	\item[$\lambda_k^{i,t}$]	Binary indicator for type 1 load. 
	\item[$\mu_k^{i,t}$]	Binary indicator for type 2 load. 
	\item[$\nu_{k,s}^{i,t},\nu_{k,e}^{i,t}$]	Integral variables to indicate starting and ending time of $k$th type 2 load. 
	\item[$\alpha_{k,n}^{i,t}$]	Element in SOS-2 $\boldsymbol{N_k^i}$ of $k$th ES. 
\end{IEEEdescription}
\subsection{Parameters}
\begin{IEEEdescription}[\IEEEusemathlabelsep\IEEEsetlabelwidth{$~~~~~~~~~$}]
	\item[${N_g}$] Number of MGs.
	\item[${T}$] Length of time horizon.
	\item[$c_b^t,c_s^t$] 	Electricity purchasing and selling price between individual MGs and the upstream grid.
	\item[$c_C^t$] 	Electricity transaction price between MGs in the community.
	\item[$\varepsilon_{ij}$]	Loss factor between MG $i$ and $j$.

	\item[$\overline{P_M^i},\underline{P_M^i}$] 	Power limits of utility grid.
	\item[$p_{C}^{i,t}$] 	Aggregation of transmitted power in community.
	
	\item[$\overline{P_C^{ij}},\underline{P_C^{ij}}$] 	Transmitted power limits. 
	\item[$\overline{P_k^{i,t}},\underline{P_k^{i,t}}$] 	ES (EV) power limits.
	\item[$\zeta_k^i$] 	ES (EV) charging/discharging efficiency.
	\item[$E_k^{i,t}$] 	ES (EV) energy level. 
	\item[$\overline{E_k^i},\underline{E_k^i}$] 	ES (EV) energy limits.
	\item[$E_k^{i,dep}$] 	EV energy requirement at departure.
	\item[$\boldsymbol{T_k^i}$]	Set of Parking time region.
	
	\item[$p_{L_1}^{i,t}$] 	Power of non-dispatchable loads.
	\item[$ p_{PV}^{i,t}$] 	Power of PV.
	\item[$p_{k}^{i,t}$] 	Power of type 1 and 2 loads. 
	\item[$P_{k}^{i}$] 	Total required energy of type 1 and 2 loads.
	\item[$H_k^i$]	Operation durations of $k$th type 1 (or 2) load sets.
	\item[$c_{k,n}^i$]	ES degradation cost coefficient.
	\item[$G_{k,n}^i$]	Energy level in which $c_{k,n}^i \in \boldsymbol{N_k^i}$ . 
	\item[$l_i^x,l_i^y$]	Coordinates of $i$th MG.
	\item[$W$]	Weighting matrix of MG community.
	\item[$w_{ij}$]	Weighting coefficient between MG $i$ and $j$.
\end{IEEEdescription}
\subsection{Functions}
\begin{IEEEdescription}[\IEEEusemathlabelsep\IEEEsetlabelwidth{$~~~~~~~~~$}]
	\item[$f_k(\bullet )$]	ES degradation cost.
	\item[$F( \bullet )$]	Linearized ES degradation cost.
\end{IEEEdescription}

\section[Introduction]{Introduction}

Microgrid (MG) is generally described as an independent small-scale power system at the downstream of the distribution system. Existing in different forms such as individual houses, commercial/residential buildings and so on \cite{kanchev2011energy, anvari2017efficient, liu2015heuristic}, MGs include a variety of distributed energy resources (DERs), e.g. photovoltaics (PVs), wind turbines (WT) and microturbines (MT), energy storages (ESs) and electric vehicles (EVs), and can operate both in the islanded mode and in conjunction with the upstream electricity grid depending on different operating requirements \cite{RN397, olivares2014trends}.

Regional MG community, such as residential area, university campus, industrial park and so on, clusters a group of MGs under the single point of common coupling (PCC) to consolidate system reliability and economy. In normal operational conditions, MG community helps individual MGs reduce transmission losses and enhance system reliability by sharing resources internally \cite{fathi2013sta}. Under extreme circumstances, MG community can operate as an autonomous entity out of the upstream power grid to maintain its integrity and security. 
Several recent studies have shown significances of MG community on power quality and system reliability improvements \cite{chiu2015mul, zhang2017scopf, 7463503}.

Energy management is the core component to improve energy efficiency, increase economic benefit and maintain operation reliability. Generally, 
energy management system (EMS) is classified into centralized and decentralized formations by different topology frameworks and control strategies to emphasize specific problems such as energy efficiency, system robustness and so on \cite{sugg_e, sugg_d, sugg_b}, in which the centralized scheme often requires full information from dispatchable components and to make decisions for each individual entity. 
However, with the size expansion of the MG community and increasing variety of power electronic components, the scheduling policy on operational strategies may become exponentially complex since intensive computation capability are often required. 
Inconsistent optimization objectives of different MGs and information security/privacy issues may also be serious concerns that hinder centralized energy management deployment into the MG community. 

Several recent studies have focused on various decentralized management schemes for MG community that require multiple entities to cooperate and coordinate in different levels, from various aspects such as cost minimization \cite{RN394, 7042324, zhang2017robust}, coalitional transaction \cite{RN389, RN392} and so forth. 
For example, a control strategy for coordinated operation is proposed in \cite{wang2015coordinated} to minimize the operational cost in a distribution system by decomposing decision-making process into multiple stages. 
A coordinated strategy for optimal energy management in multi-MG systems is presented in \cite{sugg_c} by introducing the probabilistic index for cost minimization.
Different algorithms for multi-MG coordination have been also investigated \cite{ouammi2015coordinated,sugg_a}. 

Existing studies have provided profound concepts to coordinated operation for MG community, however, they have limited consideration in several aspects as follows:
 
1) Optimization objectives of individual MGs may be inconsistent with each other as they are considered as self-interested entities. Various infrastructures of MGs would challenge the implementation of EMS in the sense of complexity on operational strategies. 

2) Existing optimization frameworks may fall into the curse of dimensionality, since they usually require intensive computation capability, especially for a large size of MGs community. Convexity issues for large-size problems may also make optimal solutions intangible \cite{papa2014decen}. 

3) Requirements for full observability in the centralized EMS and extensive information exchange in the decentralized EMS would introduce MG security and privacy issues. 

To overcome the above drawbacks, we propose a hierarchically coordinated EMS model to establish an effective and computationally efficient mechanism of power scheduling for individual MGs as well as energy transactions inside MG community. In contrast to previous existing studies, the main contributions are summarized as follows.

1) A hierarchically coordinated EMS for a regional community comprising multiple small-scale MGs is developed, aiming to minimize the total operational cost and maximize individual benefits simultaneously. 

2) In each individual MG, the local EMS is designed based on the detailed modeling of various energy resources including PVs, ESs, EVs and dispatchable loads to decide optimal power dispatches for the operational cost minimization. At the community level, a novel pairing algorithm is proposed to explicitly find appropriate MG pairings with power surplus and deficit. 
Based on the local scheduling in prior,
the community-level EMS employs the pairing algorithm to settle specific power exchanges among MGs, minimizing the total energy transactions with the upstream grid.
The individual operational cost is further reduced by additional economic benefits procured by the community-level EMS. 

3) The proposed EMS has distinguished advantages on modeling generality, computational efficiency and privacy protection: 
\\\hspace*{4mm}\textit{a)}~Each local optimization problem is formulated as a mixed-integer linear program (MILP), which can be solved by using existing free/commercial solvers in parallel; 
\\\hspace*{4mm}\textit{b)}~Computational speed is improved significantly by the non-iterative coordination strategy; and 
\\\hspace*{4mm}\textit{c)}~No information exchange among local EMSs is required, and communications between the community-level EMS and local EMSs only involve total energy exchanged with individual MGs. Therefore, private information such as detailed scheduling in MGs is well preserved.

The remainder of this paper is organized as follows. 
In Section~\ref{sec:overall}, the overall system structure is presented. 
The mathematical optimization model of individual MGs is formulated in Section \ref{sec:individual}.
Section~\ref{sec:community} elaborates the pairing algorithm in the community-level EMS and the overall coordination strategy of the MG community. 
Case studies and simulation results are discussed in Section~\ref{sec:cases}, in which the proposed EMS is simulated and compared with several benchmark approaches. 
At last, Section~\ref{sec:conclusion} summarizes the conclusion of this paper and the future work is presented.

\section{Structure of Microgrid Community} \label{sec:overall}
\begin{figure}[!tbp] 
	\centering
	\includegraphics[width=0.96\columnwidth,clip]{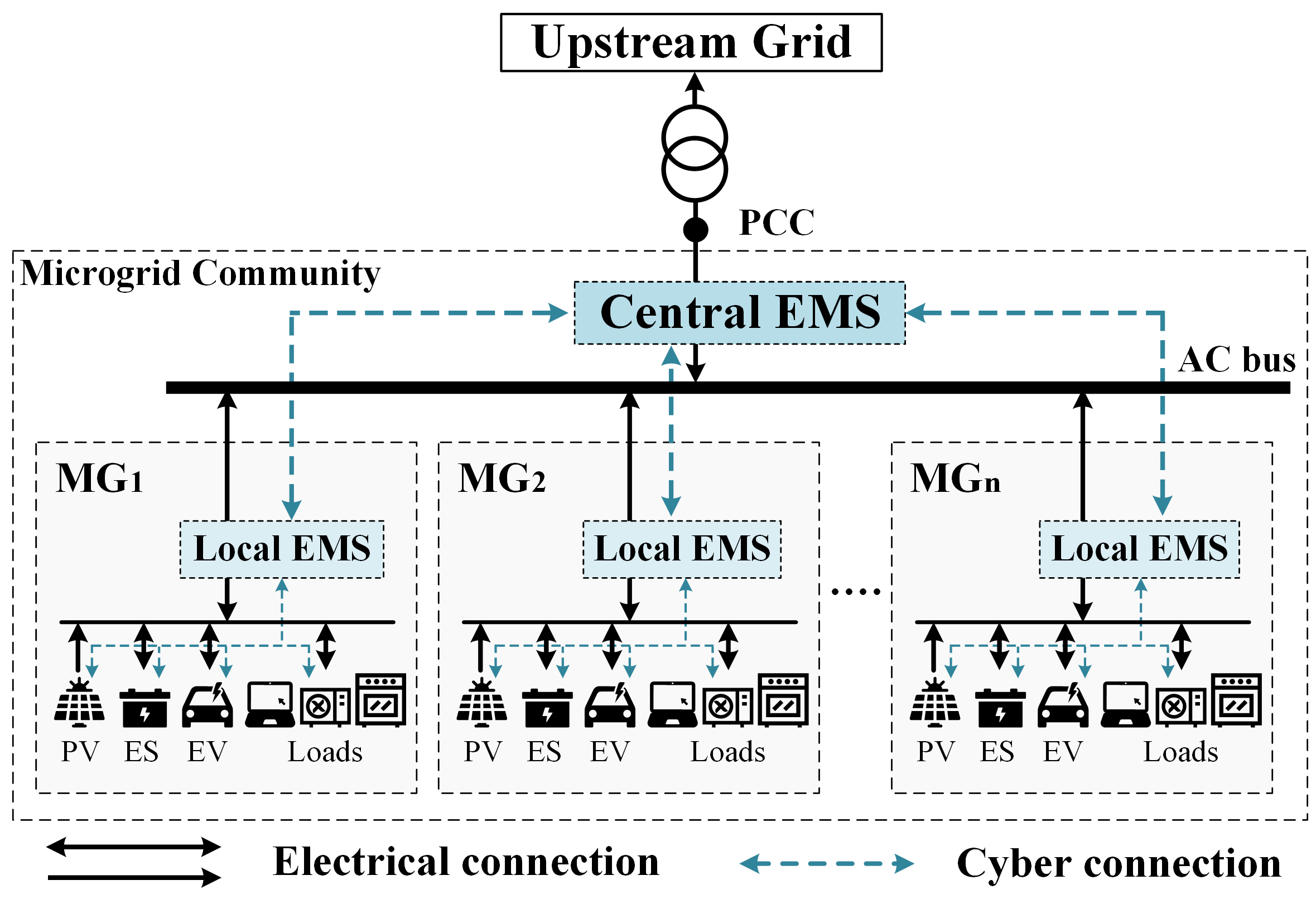}
	\par
	\vspace{-8pt}
	\caption{Network topology of a regional MG community.}
	\vspace{-8pt}
	\label{fig:schemetic}
\end{figure}

The coordinated control and communication architecture of the MG community considered in the study is schematically illustrated in Fig.~\ref{fig:schemetic}.
It is described as a small regional distributed power system comprising a community-level EMS and multiple MGs sited on different locations. The MGs are connected with the transformer to the upstream grid through a common AC bus. The point where all the MGs and the transformer have a common connection is regarded as the PCC. Each MG is locally equipped with PV, ES, EV and different types of loads. The generalized model can be easily modified and utilized by changing these components to specify different system frameworks, e.g. building clusters, residential areas, etc. 

It is seen from Fig.~\ref{fig:schemetic} that the internal operation of each individual MG are supervised by the local EMS, which only need to satisfy its self-interest. The MG-level EMSs optimize the power scheduling for local controllable units, such as minimization of the operational cost. Microgrids can also sell excess electricity back to the grid when the bidirectional transaction is allowed. 
On the other side, the community-level EMS aims to manage the operation optimally for the entire community by coordinating individual MGs, so that power efficiency and economic benefits for both individual MGs and the community are maximized simultaneously. To this extent, cyber communication is required by interaction between EMSs in different scales.

The dynamic electricity price is determined usually by the system operator from the upstream grid. It is sensitive to locational marginal prices and often announced hours ahead, allowing the decision makers to decide the power setpoints in advance. For all the MGs connected to the PCC, the same price scheme should be received since they are physically connected to the same bus. Typically, there is a difference on the purchasing and selling price at each time period to prevent economic arbitrage. 
However, concerning on the self-operation status, some MGs may sell excessive electricity back to the grid while some others may purchase to offset energy deficits. For the entire MG community, this phenomenon would lead to an undesirable result in terms of both power and economy efficiency. 
Frequent transactions between the MG community and the upstream grid would introduce additional power transmission losses; on the other hand, uneconomic operation of the MG community would become an issue due to buying and selling price differences. 

To address the problems, the proposed EMS aims to minimize energy transactions with the upstream grid such that power efficiency and economic benefits for both individual MGs and the entire community are maximized simultaneously.
Details of the proposed hierarchically coordinated EMS will be elaborated in Section~\ref{sec:individual} and \ref{sec:community}. 

\section{Local EMS in Individual Microgrid}\label{sec:individual}
As a self-interested entity, the EMS in each individual MG aims to determine the optimal scheduling for the local operational cost minimization. 
It collects all the information including cost functions and various constraints of local facilities and determines input references of the control systems within a finite scheduling time. 

\subsection{Objective Function}
Without loss of generality, the individual objective function can be formulated as follows:

\begin{align}\label{eq:objfunction}
\mathop {\min } \sum_{t \in T}{\left\{ \begin{array}{l}
	p_{M,b}^{i,t} {c_b^t} \Delta t + p_{M,s}^{i,t} {c_s^t} \Delta t \\
	+ \sum\limits_{j  \in \{ \boldsymbol{N_G}-i \} } {( p_{C,b}^{ij,t} + p_{C,s}^{ij,t} ) {c_C^t} \Delta t} \\
	+ \sum\limits_{k \in E{S^i}} {{f_k}[(p_{k,b}^{i,t} - \zeta _k^ip_{k,s}^{i,t})\Delta t]} 
	\end{array} \right\}},~ i \in \boldsymbol{N_G}
\end{align}

Each local EMS aims to minimize the operational cost within the time $\boldsymbol{T}$, which depends on its own operational requirements and is not necessarily correspondent with the community-level EMS. 
Investment costs of controllable appliances such as EV and PV are not considered, since the proposed method is focused on operational optimization where investment planning is out of this work's scope. 

The first row in \eqref{eq:objfunction} represents the electricity cost with the upstream grid. The time-varying dynamic pricing scheme provides economic incentives by bilateral transaction, in which the purchasing price $c_b^t$ is usually higher than the selling price $c_s^t$ to prevent energy arbitrage. 
The second row represents the transaction cost by power exchanges in the community, in which the transaction price $c_C^t$ is predefined as the average of  $c_b^t$ and $c_s^t$.
The last row represents the degradation cost associated with ES. It is known that the degradation process of ES is nonlinearly related with its lifetime and operation mode \cite{TR126}. To take the degradation cost in a practical fashion, the linearized model is defined from its nonlinear form in \cite{ju2017two} by using inclining blocks, as shown in Fig.~\ref{fig:deg_cost}. The ES degradation cost $f_k$ with special ordered set of type 2 (SOS-2) constraints can be written as follows: 

\begin{figure}[!tbp] 
	\centering
	\includegraphics[width=0.96\columnwidth,clip]{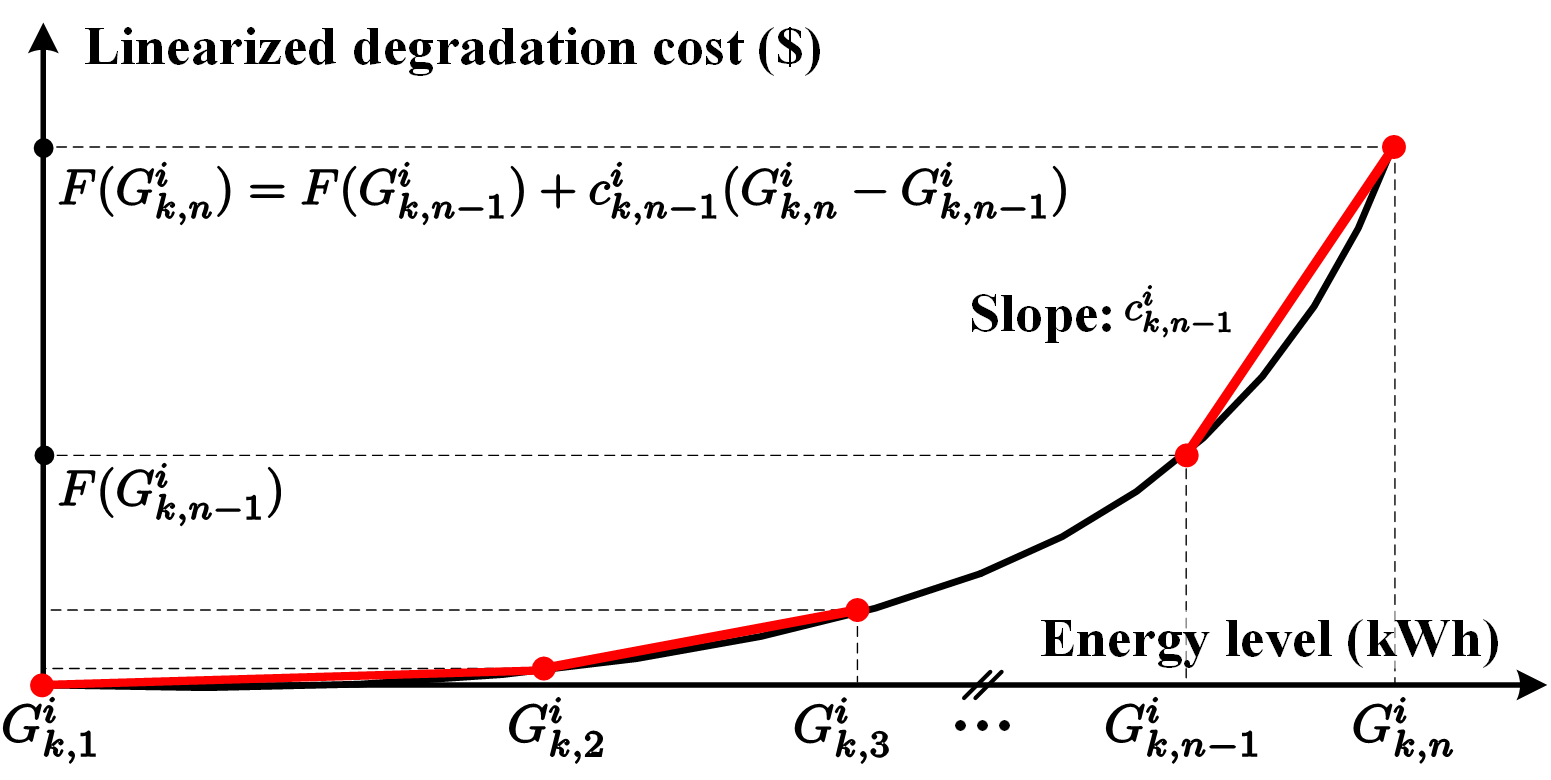}
	\par
	\vspace{-8pt}
	\caption{Piecewise linearized degradation cost of ES.}
	\label{fig:deg_cost}
\end{figure} 

\begin{align}
{f_k}(g_k^{i,t}) = \sum\limits_{n \in \boldsymbol{N_k^i}} {\alpha_{k,n}^{i,t} F(G_{k,n}^i)} \label{eq:deg_1} \\ 
g_k^{i,t} = {(p_{k,b}^{i,t} - \zeta _k^ip_{k,s}^{i,t})\Delta t} = \sum\limits_{n \in \boldsymbol{N_k^i}} {\alpha_{k,n}^{i,t} G_{k,n}^i} \\
\sum\limits_{n \in \boldsymbol{N_k^i}} {\alpha_{k,n}^{i,t} = 1},~\alpha _k^{i,t} \in [0,1] \\
F(G_{k,n}^i) {=} F(G_{k,n - 1}^i) {+} c_{k,n - 1}^i(G_{k,n}^i {-} G_{k,n - 1}^i),n \in \boldsymbol{N_k^i}{\backslash}\{1\} \\
F(G_{k,1}^i) = 0 \label{eq:es_deg_4}
\end{align}
where $\alpha_{k,n}^{i,t}$ is the element of $k$th ES in $\boldsymbol{N_k^i}$ that only the adjacent elements are nonzero.
$c_{k,n}^i$ is the degradation cost coefficient of $k$th ES, and $G_{k,n}^i$ is the power level at $c_{k,n}^i$.

\subsection{Constraints}\label{subsec:constraints}
The objective function \eqref{eq:objfunction} is subject to the various constraints on power balance, energy exchange, ES, EV and loads.

\subsubsection{Power Balance}
The total supply and demand must be always balanced which can be written as follows: 
\begin{align}
\begin{array}{r}
p_{M,b}^{i,t} + p_{M,s}^{i,t} + p_{PV}^{i,t}
+ \sum\limits_{j  \in \{ \boldsymbol{N_G}-i \} } {( p_{C,b}^{ij,t} + p_{C,s}^{ij,t} )} \\
+ \sum\limits_{k \in \boldsymbol{{ES}^i} }{( \zeta_k^i p_{k,b}^{i,t} + p_{k,s}^{i,t} )} 
+ \sum\limits_{k \in \boldsymbol{{EV}^i} }{( \zeta_k^i p_{k,b}^{i,t} +  p_{k,s}^{i,t} )} \\
 = p_{L_0}^{i,t} 
+ \sum\limits_{k \in \boldsymbol{L_1^i}} {p_k^{i,t}\lambda _k^{i,t} } 
+ \sum\limits_{k \in \boldsymbol{L_2^i}} {p_k^{i,t}\lambda _k^{i,t}} 
\end{array} \label{eq:power_balance} 
\end{align}
The left side in \eqref{eq:power_balance} includes purchasing and selling power of the upstream grid, PV power output, transmitted power with other MGs, power of ES and power of EV. The right side includes variables composing three types of loads. 

\subsubsection{Power Exchange} 
The constraints include power exchange variables to the upstream grid and to other MGs as follows, respectively: 
\begin{align}
0 \le p_{M,b}^{i,t} \le \overline{p_M^i}u_M^{i,t},~t \in \boldsymbol{T}  \label{eq:power_cont_1} \\
\underline{p_M^i}(1 - u_M^{i,t}) \le p_{M,s}^{i,t} \le 0 \label{eq:power_cont_2},~t \in \boldsymbol{T}\\
0 \le p_{C,b}^{ij,t} \le \overline{p_C^{ij}}{u_C^{ij,t}},~t \in \boldsymbol{T},j \in \boldsymbol{N_G}\backslash\{i\} \label{eq:power_cont_3}\\
\underline{p_C^{ij}}(1 - u_C^{ij,t}) \le p_{C,s}^{ij,t} \le 0,~t \in \boldsymbol{T},j \in \boldsymbol{N_G}\backslash\{i\} \label{eq:power_cont_4} 
\end{align}
where binary variables $u_M^{i,t}$ and $u_C^{ij,t}$ enforce the unidirectional power exchange at each time interval.

\subsubsection{ES and EV}
Constraints for ES are presented as follows:
\begin{align}
E_k^{i,t + 1} = E_k^{i,t} - p_{k,b}^{i,t}\Delta t - \zeta_k^i  p_{k,s}^{i,t} \Delta t,~ t \in \boldsymbol{T}, k \in \boldsymbol{{ES}^i} \label{eq:ES_dynamic}\\
\underline{E_k^i} \le E_k^{i,t} \le \overline{E_k^i},~ t \in \boldsymbol{T}, k \in \boldsymbol{{ES}^i} \label{eq:ES_cont_1}\\
0 \le p_{k,b}^{i,t} \le \overline{p_{k}^{i,t}} \delta_k^{i,t},~ t \in \boldsymbol{T}, k \in \boldsymbol{{ES}^i} \label{eq:ES_cont_2}\\
\underline {p_{k}^{i,t}} (1 - \delta_k^{i,t}) \le p_{k,s}^{i,t} \le 0,~ t \in \boldsymbol{T}, k \in \boldsymbol{{ES}^i} \label{eq:ES_cont_3} 
\end{align}
\eqref{eq:ES_dynamic}-\eqref{eq:ES_cont_3} states the state dynamic, energy limits and power limits for ES, respectively. 

Constraints for EV are similarly described as follows:
\begin{align}
E_k^{i,t + 1} = E_k^{i,t} - p_{k,b}^{i,t}\Delta t - \zeta_k^i p_{k,s}^{i,t}\Delta t,~t \in \boldsymbol{T_k^i}, k \in \boldsymbol{{EV}^i} \label{eq:EV_dynamic} \\
\underline {E_k^i}  \le E_k^{i,t} \le \overline {E_k^i} ,~t \in \boldsymbol{T_k^i}, k \in \boldsymbol{{EV}^i} \label{eq:EV_cont_1} \\
0 \le p_{k,b}^{i,t} \le \overline{p_{k}^{i,t}} \theta_k^{i,t},~t \in \boldsymbol{T_k^i}, k \in \boldsymbol{{EV}^i} \label{eq:EV_cont_2}\\
\underline {p_{k}^{i,t}} (1 - \theta_k^{i,t}) \le p_{k,s}^{i,t} \le 0,~t \in \boldsymbol{T_k^i}, k \in \boldsymbol{{EV}^i} \label{eq:EV_cont_3} \\
E_k^{i,t} \ge E_k^{i,dep},~t=T_k^i, k \in \boldsymbol{{EV}^i} \label{eq:EV_cont_4} 
\end{align}
Differently, the parking time is denoted by $\boldsymbol{T_k^i}$  that EV can be scheduled only when parked in MGs. Besides, \eqref{eq:EV_cont_4} imposes the minimum energy requirement at departure. 

\subsubsection{Loads}
Loads are classified into non-dispatchable and dispatchable loads. Non-dispatchable loads represent fixed electricity consumption that cannot be shifted over time, which are modeled as an aggregated time-dependent parameter $p_{L_0}^{i,t}$. 
Dispatchable loads represent electrical appliances which can be flexibly scheduled. Based on different operation modes, two types of dispatchable loads are defined. Type 1 loads can be dispatched to several nonconsecutive time intervals, such as washing machines that can do the wash and spin processes in different time periods. They are formulated as follows: 
\begin{align}
\sum\limits_{t \in \boldsymbol{T}} {p_k^{i,t} \lambda _k^{i,t} \Delta t}  = P_k^{i,t},~k \in \boldsymbol{L_1^i} \label{eq:L1_1}\\
\sum\limits_{t \in \boldsymbol{T}} {\lambda _k^{i,t}}  = H_k^i,~k \in \boldsymbol{L_1^i} \label{eq:L1_2} 
\end{align}
where \eqref{eq:L1_1} indicates the scheduled operation must meet total energy requirement, and \eqref{eq:L1_2} imposes the non-consecutive operational time constraint with the binary variable $\lambda_k^{i,t}$ indicating operational status and the total duration $H_k^i$.

Type 2 loads represent appliances such as toasters and dishwashers that must be scheduled consecutively. 
Their constraints are modeled as follows:
\begin{align}
\sum\limits_{t \in \boldsymbol{T}} {p_k^{i,t} \mu_k^{i,t} \Delta t}  = P_k^{i,t},~k \in \boldsymbol{L_2^i} \label{eq:L2_1}\\
\sum\limits_{t \in \boldsymbol{T}} {\mu _k^{i,t}}  = H_k^i,~k \in \boldsymbol{L_2^i} \label{eq:L2_2}\\
\sum\limits_{t \in \{\boldsymbol{T}-T\}} {\nu _{k,s}^{i,t}}  = 1,~\nu_{k,s}^{i,t} \in \left\{ {0,1} \right\},~k \in \boldsymbol{L_2^i} \label{eq:L2_3}\\
\sum\limits_{t \in \{\boldsymbol{T}-T\}} {\nu _{k,e}^{i,t}}  = -1,~\nu_{k,e}^{i,t} \in \left\{ {0, - 1} \right\},~k \in \boldsymbol{L_2^i} \label{eq:L2_4}\\
\mu _k^{i,t+1} - {\mu _k^{i,t}} = \nu _{k,s}^{i,t} + \nu _{k,e}^{i,t},~t \in \{\boldsymbol{T}-T\},~k \in \boldsymbol{L_2^i} \label{eq:L2_5}
\end{align}
where \eqref{eq:L2_3}-\eqref{eq:L2_5} are additionally imposed with integral variables $\{\nu_{k,s}^{i,t},~\nu_{k,e}^{i,t}\}$ indicating starting and ending time to address the consecutive operation feature.

\subsection{Overall Formulation} \label{sec:loc_overall}
Each individual optimization problem $\boldsymbol{M_i}$ for MG $i$ can be described as follows: 

\vspace{5pt} \noindent $\rm \boldsymbol{M_i}$: for $i \in \boldsymbol{N_G}$: 
\begin{align}
\min:~~~&\eqref{eq:objfunction} \notag \\
\mathrm{subject~to:~~~}&\eqref{eq:deg_1}-\eqref{eq:L2_5} \notag
\end{align}

$\boldsymbol{M_i}$ is formulated as a MILP which can be effectively solved by many open-source and commercial solvers. 

\section{Central EMS in Microgrid Community} \label{sec:community}
The community-level EMS aims to determine a pricing mechanism of settling transactions for MGs to further reduce their operating costs.
It allocates internal power exchanges among MGs to minimize the energy transactions with the upstream grid. 
In this section, a pairing algorithm is proposed to explicitly find the MG pairings with least power transmission distances, equivalently to minimize power losses and energy transactions. The coordination strategy is presented to determine specific power exchanges among MGs with surplus and deficit so that the transmission loss is equivalently minimized. Consequently, the individual operational cost is further reduced by additional economic benefits procured by the community-level EMS. 

\subsection{Pairing Algorithm}
The pairing algorithm identifies each distinct pair of MGs with minimal power transmission losses, so that the pattern of exchanging energy inside the MG community can be established. To achieve this target, appropriate weighting coefficients need to be addressed to determine the pairing priority. 
Since the MG community is particularly regional, electrical distances of MGs are considered to be so close that specific line resistance parameters may not be attained explicitly \cite{ding2017new, chen2016resilient}. 
Therefore, geographical distances are used instead as weighting coefficients to indirectly reflect transmission losses induced by power exchange among MGs, since they are approximately proportional to line resistances within small regions \cite{kashem2000novel, jabr2012minimum}. 
The loss factor coefficient $\varepsilon_{ij}$ is specified as well to present the linear transmission loss in a practical fashion.

Accordingly, a 2-D Cartesian coordinate system is formulated where the location of MG $i$ is expressed by its geographical coordinates $(l_i^x,l_i^y)$. Hence, the weighting coefficient $w_{ij}$ of MGs $\{i,j\}$ representing the transmission loss can be expressed by their Euclidean distance as follows:
\begin{align}
{w_{ij}} = \varepsilon_{ij} \sqrt {{{(l_i^x - l_j^x)}^2} + {{(l_i^y - l_j^y)}^2}},~i \ne j,~i,j \in \boldsymbol {N_G} \label{eq:weight_coeff}
\end{align}
Note that $w_{ij}$ for MGs with no physical connection can be set as a large positive number $M$, and intuitively, $w_{ii}=M,~i \in \boldsymbol {N_G}$  as no self connection exists. The weighting matrix $W$ of the entire community is expressed as follows:
\begin{align}
W = \left[ {\begin{array}{*{20}{c}}
	{{w_{11}}}&{{w_{12}}}& \cdots &{{w_{1{N_g}}}}\\
	{{w_{21}}}&{{w_{22}}}& \cdots &{{w_{2{N_g}}}}\\
	\vdots & \vdots & \ddots & \vdots \\
	{{w_{{N_g}1}}}&{{w_{{N_g}2}}}& \cdots &{{w_{{N_g}{N_g}}}}
	\end{array}} \right] \label{eq:weight_matrix}
\end{align}

It can be easily recognized that $W$ is ${N_g} {\times} {N_g}$ symmetric. Each row ${r_i} = \{w_{ij}|j \in \boldsymbol{N_G}\}$ in $W$ comprises the weighting coefficients of MG $i$ to other MGs. Next, we prove that there always exists a pairing $(w_{ij},w_{ji})$ in $W$ for the MG community set $\boldsymbol{N_G}$, in which they are of minimal values in their corresponding rows ${r_i}$ and ${r_j}$. 
\newtheorem{theorem}{Theorem} 
\begin{theorem}[Pairing Algorithm]\label{thm:pairing_thm}
	A MG pairing with minimal weighting coefficients to their corresponding rows in $W$ can be always found, provided that $W$ is symmetric. 
\end{theorem}
\begin{IEEEproof}[Proof of Theorem 1]
	We denote $i,j \in \boldsymbol{N_G}$ to be the indices of corresponding rows and columns in $W$, respectively. The minimum in each row $r_i$, $w_i$, can be expressed as follows:
	\begin{align}
	{w_i} = \min r_i = \mathop {\min }\limits_{j \in \boldsymbol{N_G}} {w_{ij}},~i \in \boldsymbol{N_G}
	\end{align}
	The set $\boldsymbol{R}$ including all $w_i$ can be further presented as:
	\begin{align}
	\begin{array}{l}
	\boldsymbol{R} = \{ {{w_i}|{w_i} = \mathop {\min }\limits_{j \in \boldsymbol{N_G}} w_{ij},~i \in \boldsymbol{N_G}} \} \\
	~~= \{ {{w_{ic_i}}|{w_{ic_i}} = \mathop {\min }\limits_{j \in \boldsymbol{N_G}} w_{ij},~i \in \boldsymbol{N_G}} \}
	\end{array} \label{eq:each_row}
	\end{align}
	where $\boldsymbol{C} = \{c_i|i \in \boldsymbol{N_G}\}$ is the column index set of minimum elements. 

	To prove by contradiction, we make its \textit{opposite proposition} that such a pairing does not exist for all MGs in $\boldsymbol{N_G}$. This opposite proposition implies that  $c_i \in \boldsymbol{C}$ are all different since $w_i \in \boldsymbol{R}$ are all different. 
	
	For symbol simplification, we denote the index set $\boldsymbol{K}$ as 
	\begin{align}
	\boldsymbol{K} = \{k_i \in \boldsymbol{C} | k_0 = 1, \cdots, k_i = c_{k_{i-1}},~i \in \boldsymbol{N_G} \} \label{thm:K}
	\end{align}
	Since $W$ is symmetric, in ${k_0}$th row, we have:
	\begin{align}
	w_{1{c_1}} = w_{{k_0}{c_{k_0}}} = w_{{c_{k_0}}{k_0}} = w_{{k_1}{k_0}} \label{thm:w_1}
	\end{align}
	It is known from \eqref{thm:K} that $w_{{k_1}{c_{k_1}}}$ is the minimum in the $k_1$th row and $k_0 \neq c_{k_1}$. Therefore, we have:
	\begin{align}
	w_{{k_1}{k_0}} > w_{{k_1}{c_{k_1}}} = w_{{c_{k_1}}{k_1}} = w_{{k_2}{k_1}} \label{thm:w_2}
	\end{align}
	Sequentially for $k \in \boldsymbol{K}$, the following formation must satisfy:
	\begin{align}
	w_{{k_0}{c_{k_0}}} > w_{{k_1}{c_{k_1}}} > \cdots >  w_{{k_{N_g-1}}{c_{k_{N_g-1}}}} =  w_{{k_{N_g}}{k_{N_g-1}}}
	\end{align}
	For the last element $w_{{k_{N_g}}{k_{N_g-1}}}$, we can get: 
	\begin{align}
	w_{{k_{N_g}}{k_{N_g-1}}} > w_{{k_{N_g}} {c_{k_{N_g}}}}
	\end{align}
	
	Recall that $c_i \in \boldsymbol{C}$ are all different, $k \in \boldsymbol{K} \backslash \{k_{N_g}\}$ are also all different. However, since $W$ is a ${N_g} \times {N_g}$ matrix, $\boldsymbol{K}$ must have exactly $N_g$ elements. Therefore, the following formation must satisfy: 
	\begin{align}
		\boldsymbol{K} \backslash \{k_{N_g}\} =  \boldsymbol{K} \label{eq:contract_pre}
	\end{align}
	\eqref{eq:contract_pre} means $w_{{k_{N_g}} {c_{k_{N_g}}}}$ is equal to at least one element in the minimum row set $\boldsymbol{R}$:
	\begin{align}
	w_{{k_{N_g}} {c_{k_{N_g}}}} \in R,~k_{N_g} \in \boldsymbol{K} \backslash \{k_{N_g}\} \label{eq:contract}
	\end{align}
	
	\eqref{eq:contract} is contradictory to the \textit{opposite proposition}. Hence, it is proved that such the pairing with minimal values for $W$ can be always found.
\end{IEEEproof}

By removing invalid pairings whose connections are not constructed and self-pairings in prior, it is straightforward to determine the MG pairings with minimal transmission losses in the community by using the pairing algorithm. Accordingly, after the minimal value of rows in $W$ is found, the pairing  $x ,y \in \boldsymbol{N_G}$ for $W$ can be determined by looking into the same values of these minimums as follows:
\begin{align}
\{x,y\} = \{i,j\} |\{ w_{ij}=w_{ji},~w_{ij} \in \boldsymbol{R},~i,j \in \boldsymbol{N_G} \} \label{eq:pairing_det}
\end{align} 

\subsection{Coordination Strategy in Microgrid Community}
The pairing algorithm has demonstrated MG pairings with minimal transmission loss can be always located for the entire community.
However, such a pairing is regarded valid between two MGs only with different power flow directions.
To this extent, the coordination strategy recognizes all the valid pairings of MGs with power surplus and deficits, determines specific values for corresponding energy transactions and simultaneously minimizes the total transaction loss in the MG community. 
To reach this target, total transmitted power of individual MGs with respect to the community need to be firstly distinguished. An auxiliary variable $p_{C}^{i,t}$ for $i$th MG is introduced as follows: 
\begin{align}
	p_{C}^{i,t} = p_{M,b}^{i,t} + p_{M,s}^{i,t} + \sum\limits_{j  \in \boldsymbol{N_G}\backslash\{i\}  } {( p_{C,b}^{ij,t} + p_{C,s}^{ij,t} )},~t \in \boldsymbol{T} \label{eq:power_inter_add}
\end{align}
where $p_{C}^{i,t}$ is the summed transmitted power of $i$th MG to all other MGs in $\boldsymbol{N_G}$ and the upstream grid. 

It is recognized that $p_{C}^{i,t}$ is positive when the $i$th MG has power surplus and negative with power deficit. 
Based on the power flow directions, therefore, the corresponding elements in $W$ can be excluded by the signs of $p_{C}^{i,t}$ as follows:
\begin{align}
w_{ij} = M,~{\rm{if}}~p_{C}^{i,t} \times p_{C}^{j,t} \ge 0,~i,j \in \boldsymbol{N_G} \label{eq:weight_matrix_mod} 
\end{align}
where $M$ is a large positive number. 

\begin{figure*}[!tb]
	\begin{align}
	\left\{ \begin{array}{r}
	p_{C,b}^{xy,t} = p_C^{x,t},~p_{C,s}^{xy,t} = 0,~p_{C,b}^{yx,t} = 0,~p_{C,s}^{yx,t} = -p_C^{x,t}/(1 -  w_{xy}),~{\rm if}~p_C^{x,t} > 0,~|p_C^{x,t}| < |(1 - w_{xy})p_C^{y,t}| \\
	p_{C,b}^{xy,t} =  - (1 - w_{xy})p_C^{y,t},~p_{C,s}^{xy,t} = 0,p_{C,b}^{yx,t} = 0,~p_{C,s}^{yx,t} = p_C^{y,t},~{\rm if}~p_C^{x,t} > 0,~|p_C^{x,t}| > |(1 - w_{xy})p_C^{y,t}| \\
	p_{C,b}^{xy,t} = 0,~p_{C,s}^{xy,t} = p_C^{x,t},~p_{C,b}^{yx,t} =  - (1 - w_{xy})p_C^{x,t},~p_{C,s}^{yx,t} = 0,~{\rm if}~p_C^{x,t} < 0,~|(1 - w_{xy})p_C^{x,t}| < |p_C^{y,t}| \\
	p_{C,b}^{xy,t} = 0,~p_{C,s}^{xy,t} =  - p_C^{y,t}/(1 - w_{xy}),~p_{C,b}^{yx,t} = p_C^{y,t},~p_{C,s}^{yx,t} = 0,~{\rm if}~p_C^{x,t} < 0,~|(1 - w_{xy})p_C^{x,t}| > |p_C^{y,t}|
	\end{array} \right. ,~t \in \boldsymbol{T} \label{eq:update_pij} \\
	\left\{ \begin{array}{r}
	p_C^{x,t} = 0,~p_C^{y,t} = p_C^{y,t} + p_C^{x,t}/(1 - w_{xy}),~{\rm if}~ p_C^{x,t} > 0,~|p_C^{x,t}| < |(1 - w_{xy})p_C^{y,t}| \\
	p_C^{y,t} = p_C^{x,t} + (1 - w_{xy})p_C^{y,t},~p_C^{x,t} = 0,~{\rm if}~ p_C^{x,t} > 0,~|p_C^{x,t}| > |(1 - w_{xy})p_C^{y,t}| \\
	p_C^{x,t} = 0,~p_C^{y,t} = p_C^{y,t} + (1 - w_{xy})p_C^{x,t},~{\rm if}~ p_C^{x,t} < 0,~|(1 - w_{xy})p_C^{x,t}| < |p_C^{y,t}| \\
	p_C^{y,t} = p_C^{x,t} + p_C^{y,t}/(1 - w_{xy}),~p_C^{x,t} = 0,~{\rm if}~ p_C^{x,t} < 0,~|(1 - w_{xy})p_C^{x,t}| > |p_C^{y,t}|
	\end{array} \right. ,~t \in \boldsymbol{T} \label{eq:pc}
	\end{align}
	\hrulefill 
\end{figure*}

After elimination of irrelevant parameters, valid pairings of MGs with different power flow directions can be always established by using the modified coefficient matrix $W$ in \eqref{eq:weight_coeff}, \eqref{eq:weight_matrix} and \eqref{eq:weight_matrix_mod}. Based on the optimized dispatch signals transmitted from local EMSs, the community-level EMS can determine the transactions among MGs explicitly. 

The community-level EMS executes the pairing algorithm to find MGs in a pairing with minimal weighting coefficients, marked as $x ,y \in \boldsymbol{N_G}$.
Correspondingly, the variables related to exchanged power of MGs $x,y$ need to be updated in the MG level. For all possible scenarios, the transmitted power of MGs $x,y$ are defined by \eqref{eq:update_pij}, and variables $p_{C}^{x,t}$ and $p_{C}^{y,t}$ indicating summed transmitted power of MGs $x,y$ respectively are updated by \eqref{eq:pc} as well.

At last, the weighting matrix $W$ is updated as follows so that the successfully paired MGs have been excluded:
\begin{align}
w_{ij} = M~\forall w_{ij} \in {r_i}~{\rm \&}~j \in \boldsymbol{N_G},~{\rm if}~p_{C}^{i,t}=0,~i \in \{x,y\} \label{eq:weight_W_exc}
\end{align}

\begin{figure}[!tbp]
	\small
	\begin{algorithmic}[1]
		\hrule\vspace{0.25em}
		\FOR {$t \in \boldsymbol{T}$, } 
		\STATE \textit{\textbf{Lower-level EMS for individual MGs:}}
		\FOR {MG $i \in \boldsymbol{N_G}$, }
		\STATE 1. Make local PV, load and electricity price forecast in $\boldsymbol{T}$. 
		\STATE 2. Solve the local optimization problem $\boldsymbol{M_i}$. 
		\STATE 3. Determine the decision variables $\{ p_{k,b}^{i,t}, p_{k,s}^{i,t}, u_M^{i,t}, \delta_k^{i,t}, \theta_k^{i,t},$ $ \lambda_k^{i,t}, \mu_k^{i,t}, \nu_{k,s}^{i,t}, \nu_{k,e}^{i,t}, \alpha_{k,n}^{i,t} \} $ and  $\{ u_M^{i,t}, u_C^{ij,t} \}$. 
		\STATE 4. Aggregate $p_c^{i,t}$ according to \eqref{eq:power_inter_add}, and transfer $\{ p_{M,b}^{i,t},$ $p_{M,s}^{i,t}, p_{C,b}^{ij,t}, p_{C,s}^{ij,t} \}$ to be decided by the community-level EMS.
		\ENDFOR
		\STATE \textit{\textbf{Community-level EMS for MG community:}}
		\STATE 1) Formulate the weighting matrix $W$ according to \eqref{eq:weight_coeff}, \eqref{eq:weight_matrix} and \eqref{eq:weight_matrix_mod}.
		\WHILE {\eqref{eq:end_cri} is met,}
		\STATE 2) Find the minimum of each row in $W$, and determine the specific pairing $\{x,y\}$ with the same smallest coefficients according to \textbf{Theorem \ref{thm:pairing_thm}} and \eqref{eq:pairing_det}. 
		\STATE 3) Update $\{ p_{c,b}^{ij,t},~p_{c,s}^{ij,t} \},~i,~j {\in} \boldsymbol{N_G} $ by \eqref{eq:update_pij}.
		\STATE 4) Update aggregated power variables $p_{C}^{i,t}\!,i\!\in\!\boldsymbol{N_G}$ by \eqref{eq:pc}.
		\STATE 5) Update the weighting matrix $W$ by \eqref{eq:weight_W_exc}.
		\ENDWHILE
		\STATE 6) Determine the power exchange variables $\{ p_{M,b}^{i,t}, p_{M,s}^{i,t}, p_{C,b}^{ij,t},$ $p_{C,s}^{ij,t} \}, ~i \in \boldsymbol{N_G}$.
		\ENDFOR
		\vspace{0.25em}
		\hrule
	\end{algorithmic}
	\caption{Algorithm of the hierarchically coordinated EMS.} 
	\vspace{-6pt}
	\label{fig:overall_algo}
\end{figure}

\eqref{eq:weight_W_exc} finishes the first MG pairing, and the community-level EMS starts searching next pairing for the rest of MGs, until any of the total energy surplus of deficit among $\boldsymbol{N_G}$ becomes zero. The stopping criterion can be expressed as follows:

\begin{align}
\sum\limits_{i \in \boldsymbol{N_G}} {\sum\limits_{j \in \{ \boldsymbol{N_G}- i\} } {p_{C,b}^{ij,t}} } \times \sum\limits_{i \in \boldsymbol{N_G}} {\sum\limits_{j \in \{ \boldsymbol{N_G}- i\} } {p_{C,s}^{ij,t}} } =0 \label{eq:end_cri}
\end{align}

The overall procedure of the proposed EMS is described in Fig.~\ref{fig:overall_algo}. 
In the beginning of each time period, PV and load forecasts are made locally and the day-ahead electricity price is obtained.  
For each MG $i \in \boldsymbol{N_G}$, dispatch decisions in the entire scheduling horizon are obtained by solving its own optimization problem in the local EMS individually, and $p_{C}^{i,t}$ is calculated by \eqref{eq:power_inter_add}. 
Then for the community-level EMS, $W$ is re-established based on \eqref{eq:weight_coeff}, \eqref{eq:weight_matrix} and \eqref{eq:weight_matrix_mod}.
Afterwards, the community-level EMS finds the MG pairing for current $W$. Procedures of the coordination strategy in the community-level EMS are iterated by \eqref{eq:power_inter_add}--\eqref{eq:weight_W_exc} until \eqref{eq:end_cri} is met. 

\subsection{Remarks}
It is worth mentioning that the proposed pairing algorithm is not to physically control the power flow from one MG to the other. Rather, it is to settle the energy transactions within different MGs in a hierarchical way. 
Specifically, the local EMS determines the optimal power dispatch for each individual MG, and all the excessive power will go to the PCC for power exchange.
The pairings of power surplus and deficit are then established based on the in-prior local scheduling in the community-level EMS. 
In this way, each MG can transact with others rather than the upstream grid to save the operating cost. 
It is important to note that the community-level EMS is not to physically dispatch the power but to settle the energy transactions, and the power surplus and deficit of all the individual MGs are balanced with the upstream grid at the PCC.
Also note in Fig.~\ref{fig:overall_algo} that the community-level EMS has no iterative information exchange with the local EMS since $p_{C}^{i,t}$ is only collected once from the corresponding $i$th local EMS, and that 
the local EMS can solve the MILP optimization problem in parallel.
Such the non-iterative interaction and parallel optimization will improve the computational speed of the proposed EMS over the existing techniques. The verification on computation speed will be provided in Section~\ref{sec:cases}.

On the selection of weighting coefficients, 
as the electrical distances of MGs are close due to the regionally small area, it is usually not feasible to obtain exact line parameters and thus hard to calculate the explicit power losses.
It is similarly suggested in \cite{turitsyn2011options, jabr2018linear} that the fixed R/X ratio can be employed approximately in the homogeneous distribution system to calculate the line parameters. 
On the other hand, since it is only required in \textbf{Theorem 1} that the weighting coefficient matrix $W$ is symmetric, the specific line parameters can be readily employed as weighting coefficients for the proposed algorithm without affecting its effectiveness if they are explicitly known beforehand.  

\section{Simulation Results} \label{sec:cases}
In this section, the mathematical model of the proposed EMS is demonstrated in MATLAB. The optimization problem is solved using Gurobi \cite{rgurobi}, in which the local EMS in individual MG solves its own optimization. The community-level EMS settles the power exchanges among MGs by using the coordination strategy with pairing algorithm, so that the total power transactions with the upstream grid is minimized. The simulation is conducted for a 24h scheduling horizon, and the sampling time resolution for the community-level EMS is 0.5h. 

\begin{table}[!tbp]
\caption{Microgrid Characteristics}
\label{tab:mg_chars}
\vspace{-8pt}
\resizebox{0.98\columnwidth}{!}{
	\begin{tabular}{l|c|c|c|c}
		\hline
		\textbf{Microgrid} & MG1& MG2 & MG3 & MG4 \\  \hline
		\textbf{ES} \\ \hline
		Capacity(kWh)&	8&	8&	12&	12 \\ \hline
		Max\&Min $P$(kW)&	-4/4&	-4/4&	-4/4&	-4/4 \\ \hline
		Initial SOC(\%)&	20.9&	33.1&	33&	31 \\ \hline
		SOC range(\%)&	17.0-84.1&	17.5-83.5&	16.9-82.1&	18.7-89.0 \\ \hline
		Efficiency&	95\%&	95\%&	95\%&	95\% \\ \hline
		\textbf{EV} \\ \hline 
		Capacity (kWh)&	16&	16&	N.A.&	N.A. \\ \hline
		Max\&Min $P$ (kW)&	-1.44/3.6&	-1.44/3.6& {}& {} \\ \hline 
		Initial SOC(\%)&	52.63&	33.1	& {}& {} \\ \hline 
		Operation periods (h)&	0-4.88, 19.09-24&	0-7.65,18.93-24	& {}& {} \\ \hline 
		SOC range(\%) (\%)&	15.8-83.7&	19.9-81.6& {}& {} \\ \hline 
		Min depart SOC(\%)&	51.45&	61.58& {}& {} \\ \hline 
		Efficiency&	95\%&	95\%& {}& {} \\ \hline
		\textbf{PV} \\ \hline
		Capacity (kWp)&	2&	2&	16&	16 \\\hline
		\textbf{Unified location} \\ \hline
		{($l^x,l^y$)}&	{(0.12, 0.13)}&	{(0.16, 0.79)}&	{(0.83, 0.11)}&	{(0.09, 0.26)} \\
		\hline
	\end{tabular}
}
\vspace{-6pt}
\end{table}

\begin{table}[!tbp]
	\tiny
	\centering
	\caption{Parameters of dispatchable loads}
	\label{tab:dispatchable_loads}
	\vspace{-8pt}
	\resizebox{\columnwidth}{!}{
		\begin{tabular}{c|c|c|c|c}
			\hline
			\multirow{2}{*}{Appliance}&\multirow{2}{*}{Power(kW)}&{Operating}&{Operation}&\multirow{2}{*}{type} \\
			{}&{}&{periods}&{duration(h)} \\\hline
			{Washing machine}&{0.7}&{0-19,23-24}&{1}&{1} \\\hline
			{Cleaner}&{0.6}&{0-4,6,24}&{4}&{1} \\\hline
			{Air conditioner}&{1.2}&{0-7,18-24}&{3}&{1} \\\hline
			{Lighting}&{0.15}&{6-7,18-23.5}&{5}&{1} \\\hline
			{Oven}&{1.16}&{11-13}&{0.5}&{1} \\\hline
			{Toaster}&{1.2}&{7-9}&{0.25}&{2} \\\hline
			{Dish washer}&{1}&{0-4,9-11,14-17,20-24}&{1}&{2} \\\hline
		\end{tabular}
	}
\vspace{-6pt}
\end{table}

\begin{figure}[!tbp]
	\centering
	\subfloat[MG1]{
		\includegraphics[width=0.48\columnwidth]{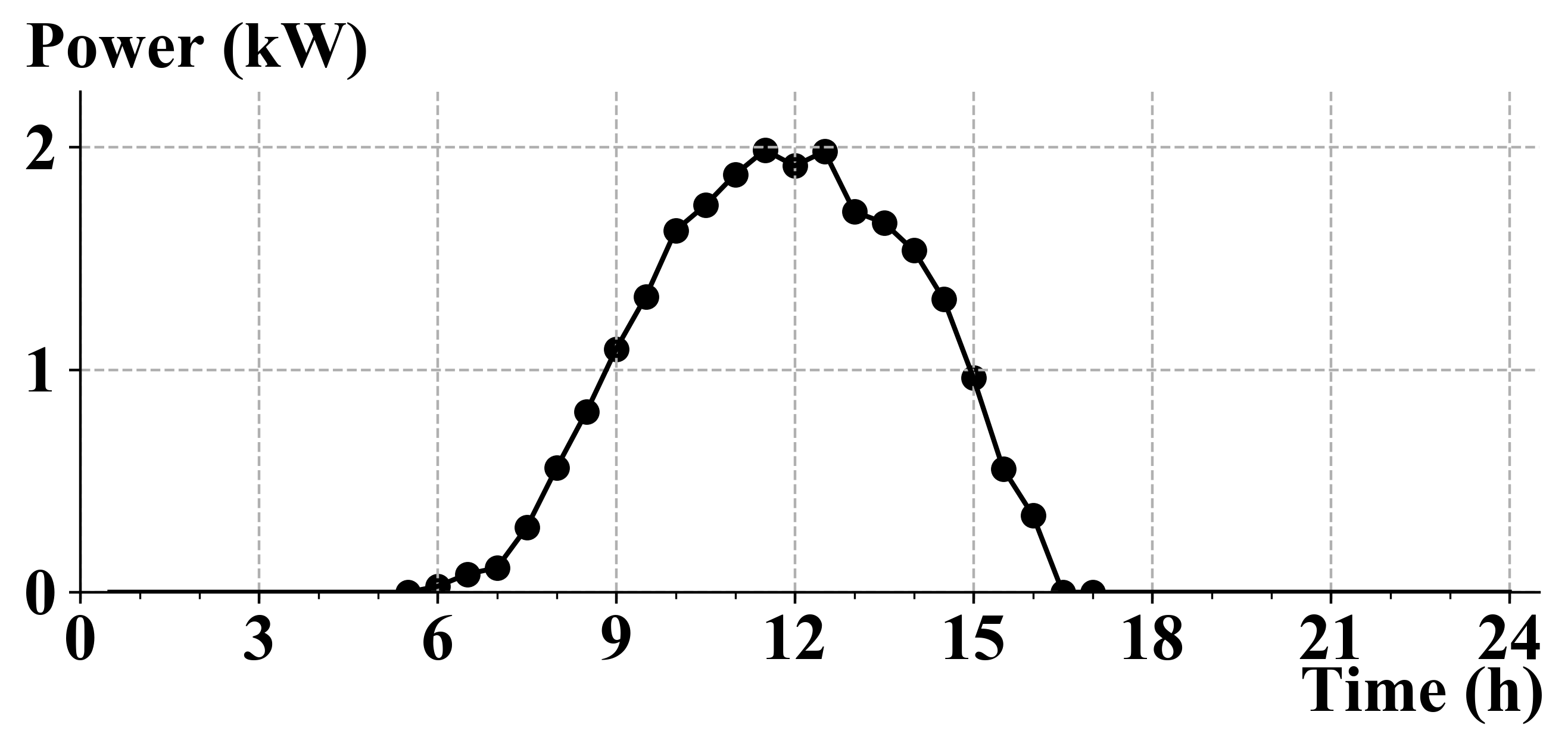}
		\label{fig:pv_profile_1}
	} 
	\subfloat[MG2]{
		\includegraphics[width=0.48\columnwidth]{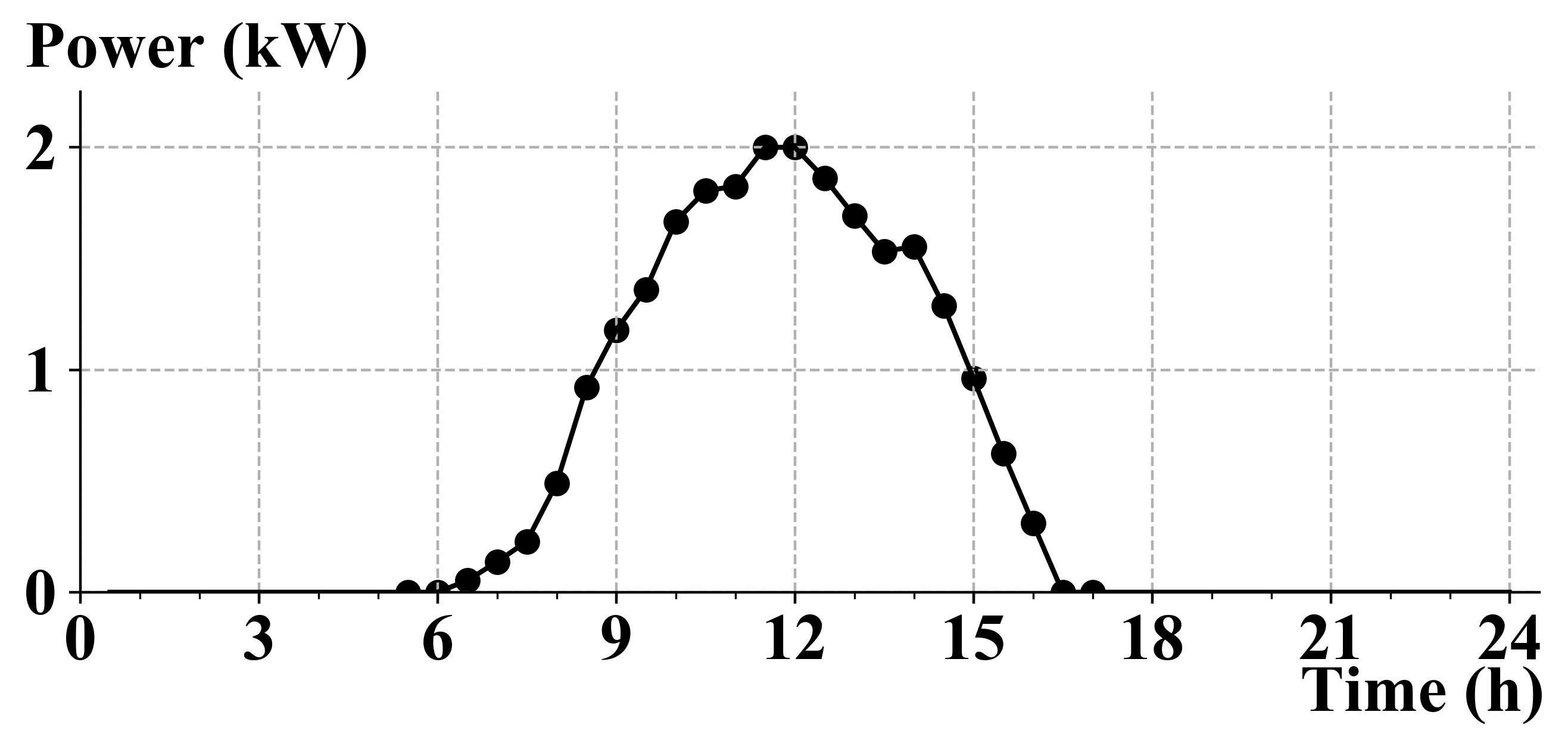}
		\label{fig:pv_profile_2}
	} \vspace{-6pt} \par
	\subfloat[MG3]{
		\includegraphics[width=0.48\columnwidth]{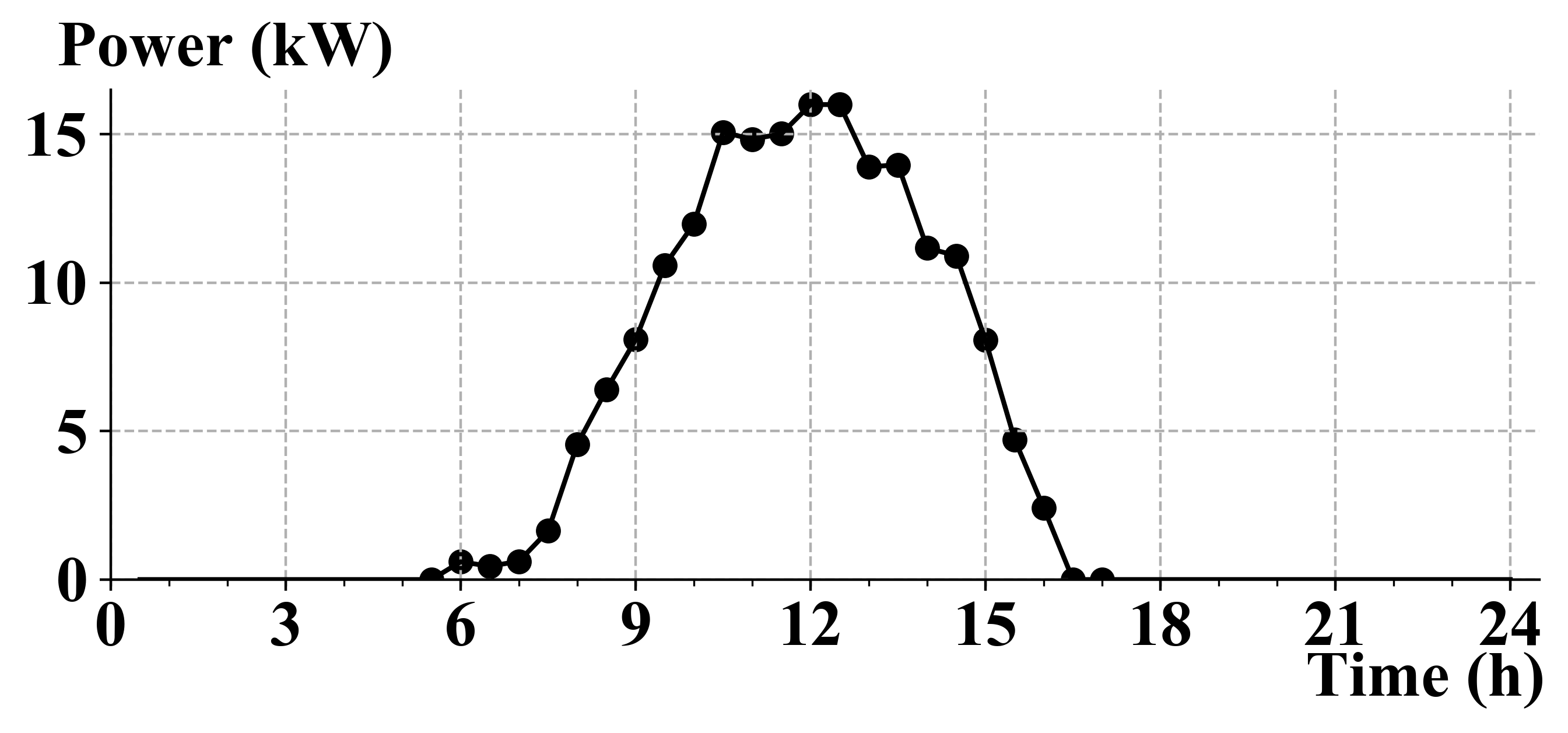}
		\label{fig:pv_profile_3}
	} 
	\subfloat[MG4]{
		\includegraphics[width=0.48\columnwidth]{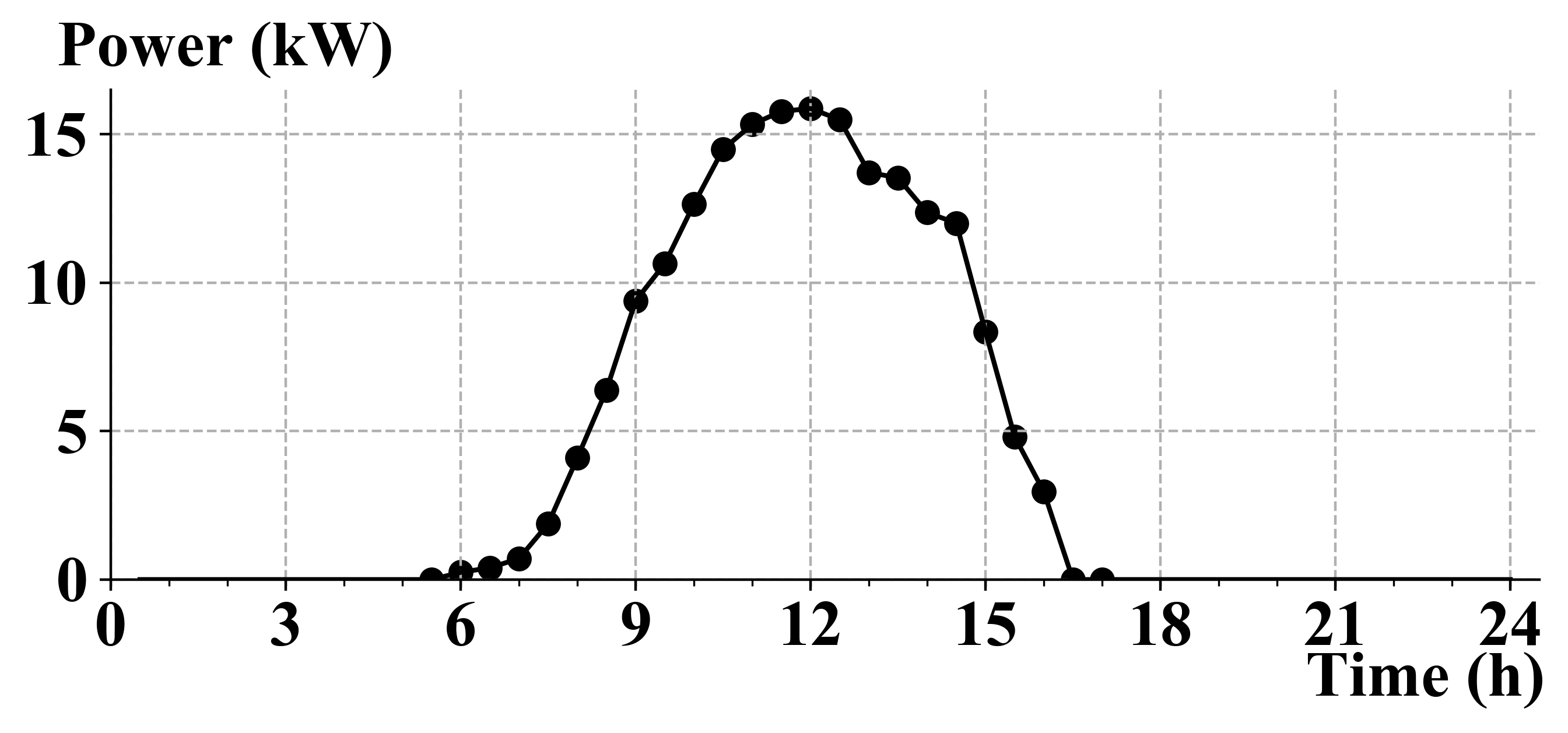}
		\label{fig:pv_profile_4}
	} \par
	\caption{PV outputs of MGs.}
	\label{fig:pv_profile}
\end{figure}

\begin{figure}[!tbp]
	\centering
	\subfloat[MG1]{
		\includegraphics[width=0.48\columnwidth]{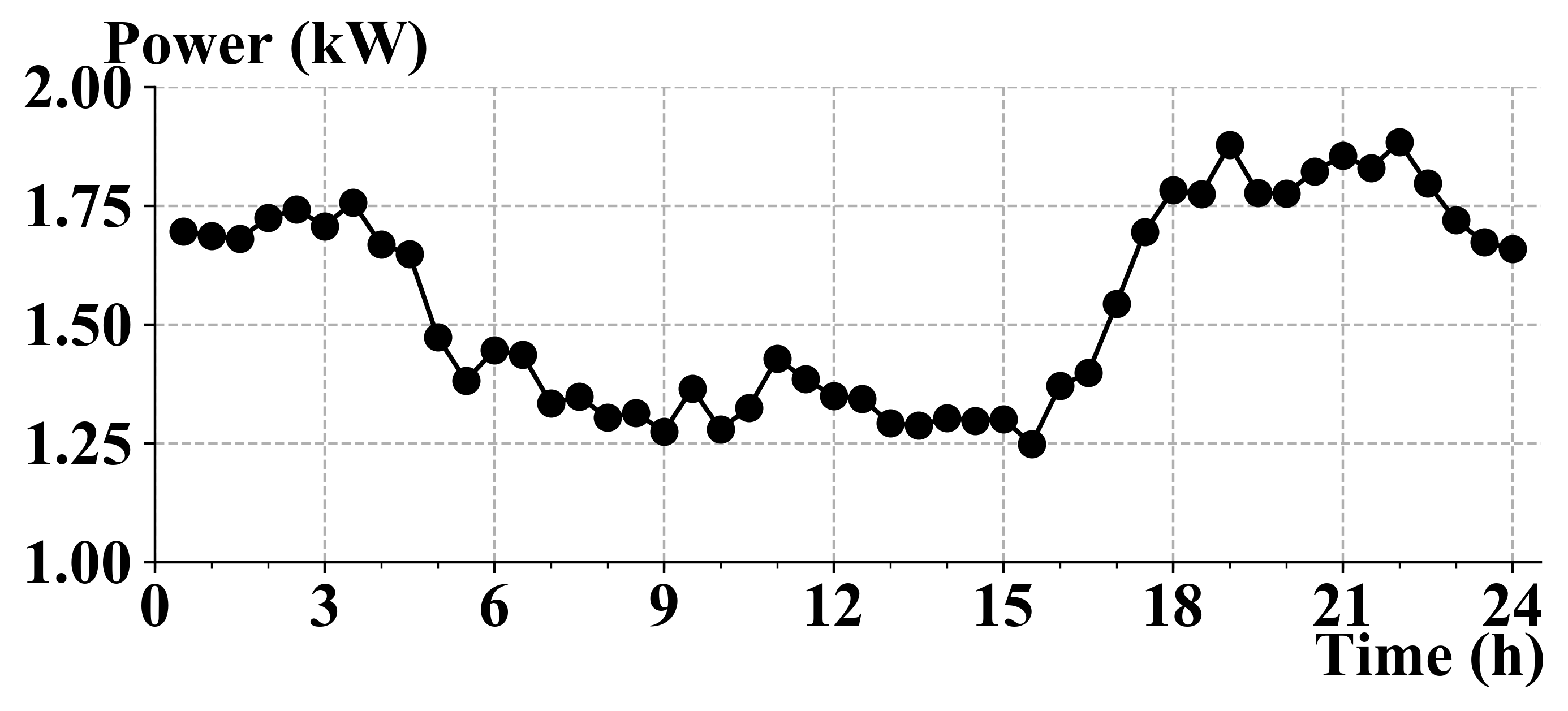}
		\label{fig:load_profile_0}
	} 
	\subfloat[MG2]{
		\includegraphics[width=0.48\columnwidth]{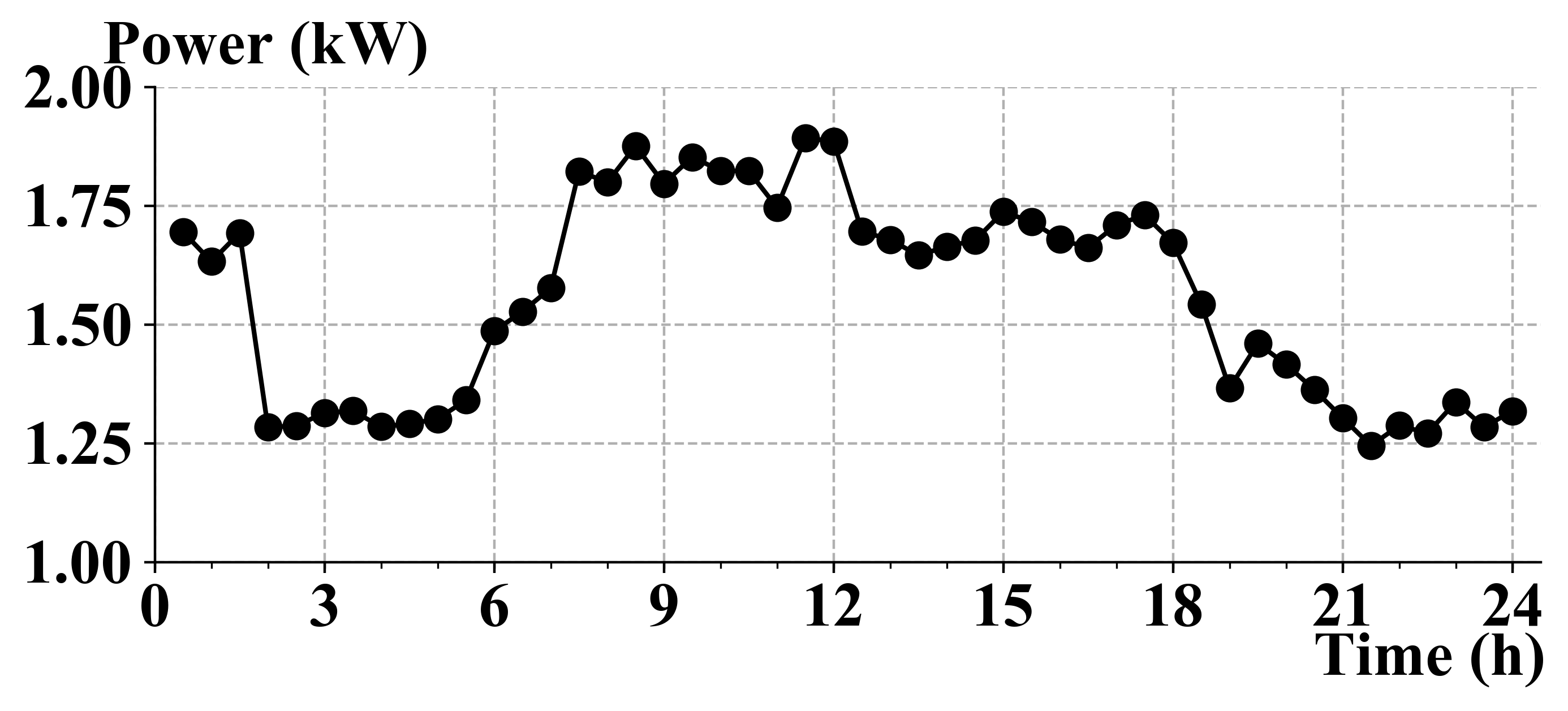}
		\label{fig:load_profile_1}
	} \vspace{-6pt} \par
	\subfloat[MG3]{
		\includegraphics[width=0.48\columnwidth]{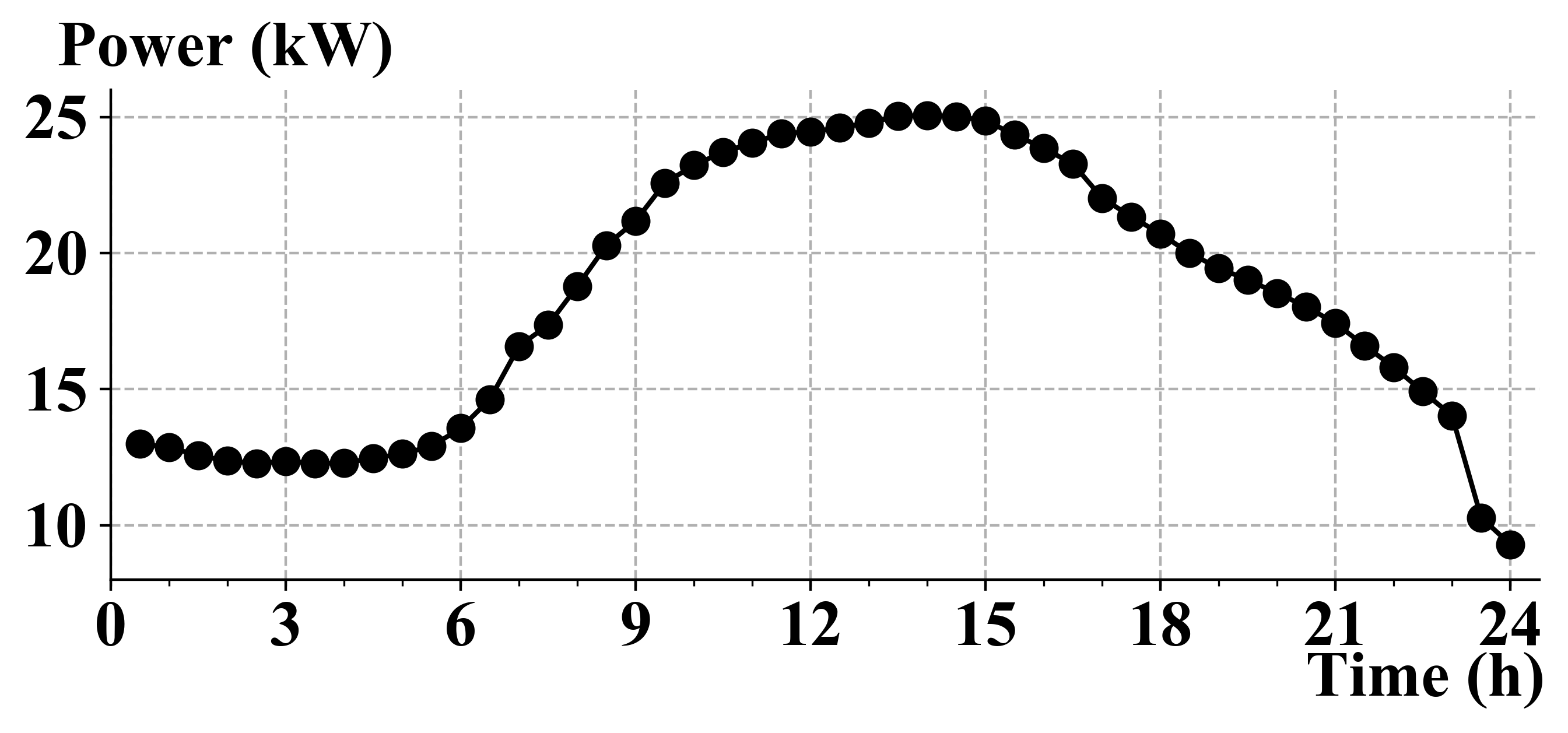}
		\label{fig:load_profile_2}
	} 
	\subfloat[MG4]{
		\includegraphics[width=0.48\columnwidth]{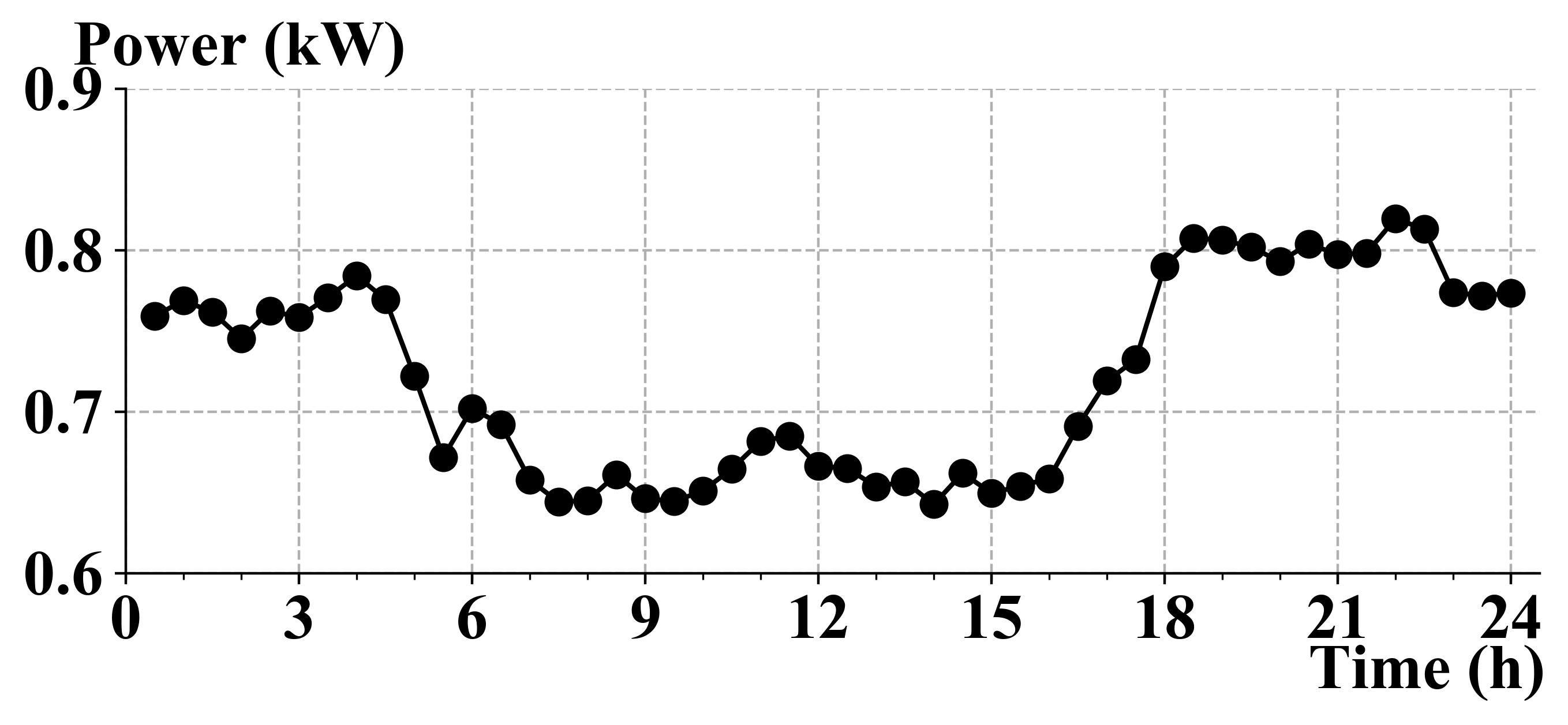}
		\label{fig:load_profile_3}
	} \par
	\caption{Non-dispatchable loads of MGs.}
	\label{fig:load_profile}
\end{figure}

\subsection{Case 1: 4-MG Community}
\begin{figure}[!tbp]
	\centering
	\includegraphics[width=0.96\columnwidth]{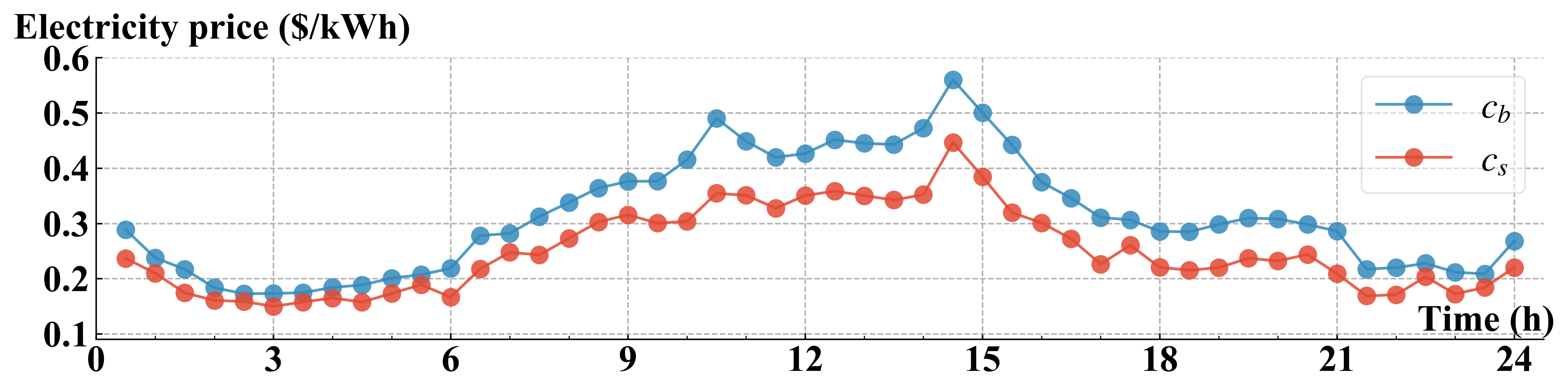}
	\par
	\vspace{-6pt}
	\caption{Electricity price in 24 hours.}
	\label{fig:price_table}
\end{figure}

In this case study, a MG community with 4 different types of MGs (MG1-MG4) is investigated.
MG1 and MG2 are considered as two individual houses with different load patterns. MG3 is an apartment building with 10 households. MG4 is a small-scale MG with high PV penetration, where its renewable output is much larger than the total load.
The specifications of 4 MGs are detailed in Table~\ref{tab:mg_chars}. For each households in individual houses and apartment, type 1 and type 2 loads are listed in Table~\ref{tab:dispatchable_loads} with predefined operation time ranges. 
The PV profile is based on the solar radiation data from \cite{SERIS}, which is depicted in Fig.~\ref{fig:pv_profile}.
The non-dispatchable load and electricity price data are based on the average household statistics in 2017 \cite{EMASG} and the hourly market data in Singapore \cite{RN207}, which are shown in Fig.~\ref{fig:load_profile} and Fig.~\ref{fig:price_table}, respectively.
Specifically, to indicate different patterns, the load profile of MG2 is shifted so that it has the peak power consumption in the daytime.
%
The loss factors in the community $\varepsilon_{ij},~i,j \in \boldsymbol{N_G}$ are set to be 0.05.

\subsubsection{Results of Individual Microgrids}
\begin{figure}[!tbp]
	\centering
	\subfloat[MG1]{
		\includegraphics[width=0.99\columnwidth]{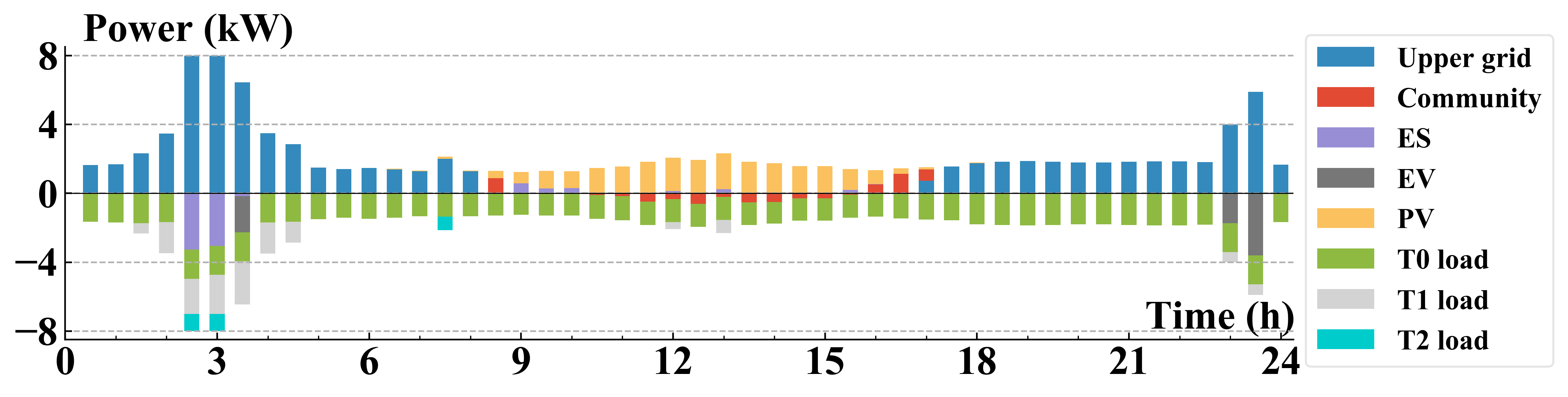}
		\label{fig:case1}
	} \vspace{-2pt} \par 
	\subfloat[MG2]{
		\includegraphics[width=0.99\columnwidth]{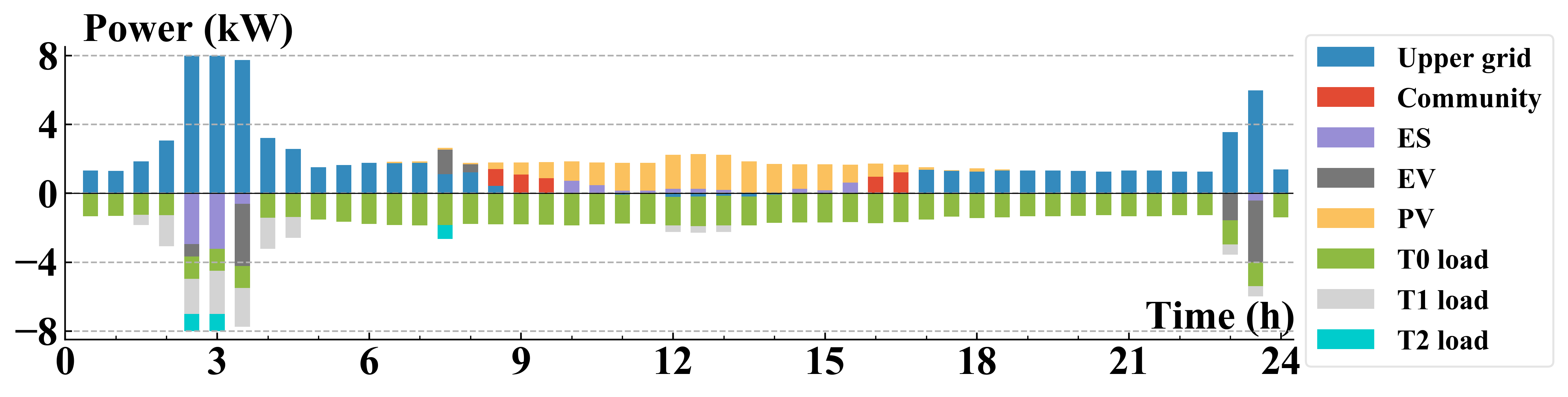}
		\label{fig:case2}
	} \vspace{-2pt} \par
	\subfloat[MG3]{
		\includegraphics[width=0.99\columnwidth]{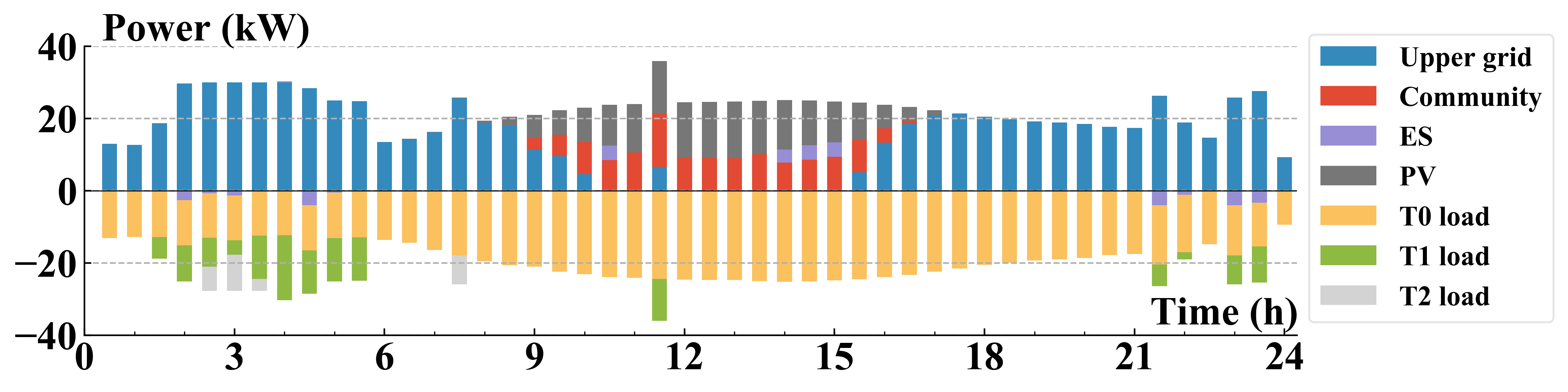}
		\label{fig:case3}
	} \vspace{-2pt} \par
	\subfloat[MG4]{
		\includegraphics[width=0.99\columnwidth]{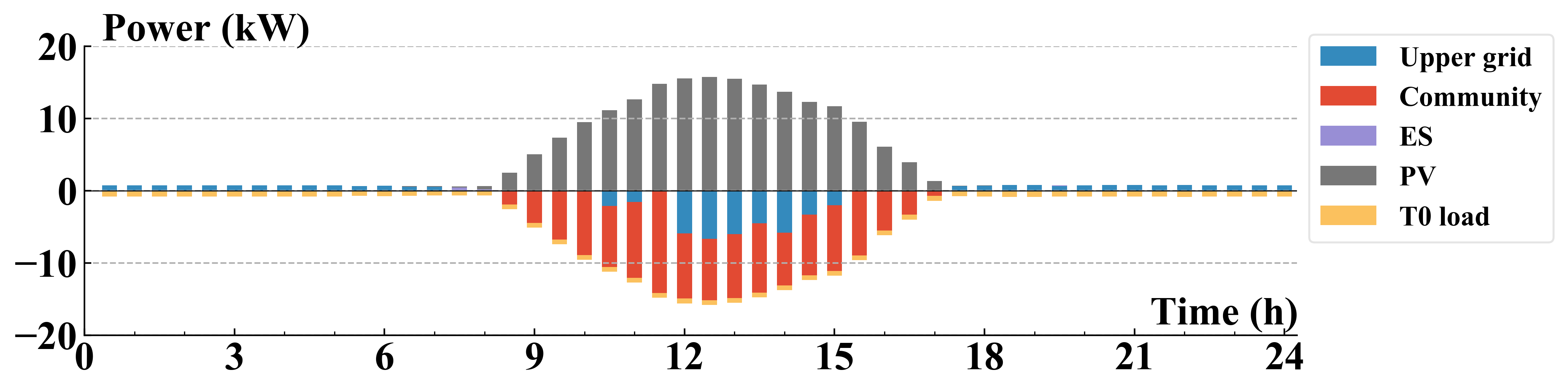}
		\label{fig:case4}
	} \par
	\caption{Energy scheduling in different MGs.}
	\label{fig:result}
\end{figure}

The hourly energy dispatch for MG1 and MG2 (houses) is presented in Fig.~\ref{fig:case1} and Fig.~\ref{fig:case2}. It is observed that MG1 and MG2 can cover most of electricity demand themselves by PV during daytime. Dispatchable loads are scheduled within off-peak hours due to low electricity price. ES and EV are also charged until the required energy levels are reached at hours with low electricity prices (e.g., at hour 2 and 23). However, ES can effectively respond high price signals in the daytime whereas EV is not involved as it is already departed. Particularly, when the electricity price increases at hour 7, EV in MG2 starts to charge to reduce electricity consumption of the upstream grid. 

\begin{figure}[!tbp] 
	\centering
	\includegraphics[width=0.99\columnwidth]{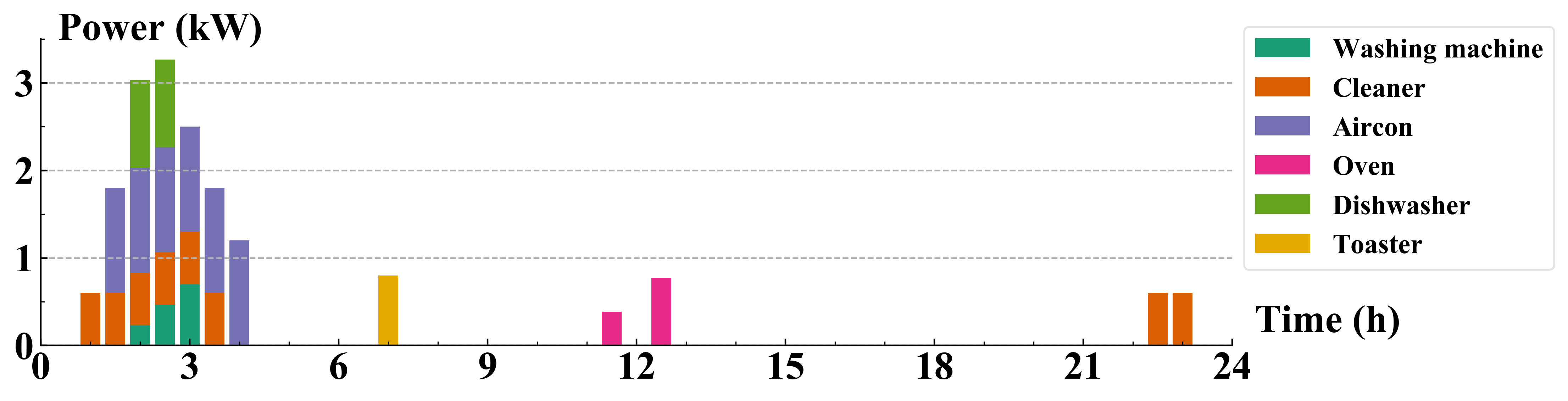}
	\par
	\vspace{-2pt}
	\caption{Detailed scheduling of dispatchable loads in MG1.}
	\label{fig:load_house1}
\end{figure}

The scheduling of appliances in MG1 is presented in Fig.~\ref{fig:load_house1} for detailed illustration. Most power consumption is distributed among time intervals when electricity prices are at lowest even full scheduling flexibility is provided. On the other side, power consumption of type 2 loads is moved towards hours with higher prices to meet users' demands. For both types of dispatchable loads, time intervals with lowest prices are always selected by the local EMS. 

The scheduling for MG3 (apartment building) is shown in Fig.~\ref{fig:case3}. Similarly, the load consumption is mostly distributed among time intervals with low electricity price. Both types of dispatchable loads are scheduled within off-peak hours even though type 1 loads have full flexibility during the entire horizon. However, different from MG1 and MG2 (houses), it is necessary for MG3 to buy part of additional power from other MGs with surplus or even from the upstream grid since the demand cannot be fully covered by its own PV output.

The scheduling of MG4 is shown in Fig.~\ref{fig:case4}, in which this MG owns small non-dispatchable loads. In consequence, the net power is exported to other MGs. On the other hand, power consumption at night is supplied by the local energy storage, the upstream grid and other MGs together since PV is unable to provide any power.

\subsubsection{Results of Microgrid community}

\begin{table}[tbp]
	\tiny
	\caption{Energy flow results in 24 hours}
	\label{tab:energy_flow}
	\vspace{-4pt} 
	\resizebox{\columnwidth}{!}{
		\begin{tabular}{c|c|c|c|c|c|c}
			\hline
			{\textbf{Time}}&\multicolumn{6}{c}{\textbf{Energy flow (kWh)}} \\ \cline{2-7}
			{\textbf{(h)}}& {1-2}&{1-3}&{1-4}&{2-3}&{2-4}&{3-4} \\ \hline
			1&	0&	0&	0&	0&	0&	0 \\ \hline
			2&	0&	0&	0&	0&	0&	0 \\ \hline
			3&	0&	0&	0&	0&	0&	0 \\ \hline
			4&	0&	0&	0&	0&	0&	0 \\ \hline
			5&	0&	0&	0&	0&	0&	0 \\ \hline
			6&	0&	0&	0&	0&	0&	0 \\ \hline
			7&	0&	0&	0&	0&	0&	0 \\ \hline
			8&	0&	0&	0&	0&	0&	0 \\ \hline
			9&	0&	0&	0&	0&	-1.0798&	-3.3395 \\ \hline
			10&	0&	0&	0&	0&	0&	-8.8526 \\ \hline
			11&	0&	0.1501&	0&	0&	0&	-10.4804 \\ \hline
			12&	0&	0.3298&	0&	0&	0&	-9.0236 \\ \hline
			13&	0&	0.2102&	0&	0&	0&	-8.8724 \\ \hline
			14&	0&	0.5107&	0&	0&	0&	-7.2907 \\ \hline
			15&	0&	0.3018&	0&	0&	0&	-9.1386 \\ \hline
			16&	0&	0&	-0.5124&	0&	-0.9727&	-3.9986 \\ \hline
			17&	0&	0&	-0.6501&	0&	0&	0 \\ \hline
			18&	0&	0&	0&	0&	0&	0 \\ \hline
			19&	0&	0&	0&	0&	0&	0 \\ \hline
			20&	0&	0&	0&	0&	0&	0 \\  \hline
			21&	0&	0&	0&	0&	0&	0 \\ \hline
			22&	0&	0&	0&	0&	0&	0 \\ \hline
			23&	0&	0&	0&	0&	0&	0 \\ \hline
			24&	0&	0&	0&	0&	0&	0 \\ \hline
		\end{tabular}
	}
\end{table}

The energy flow within the MG community is detailed in Table~\ref{tab:energy_flow}, in which four MGs are abbreviated from 1 to 4, respectively. It is shown that MG1 has the higher priority to export its surplus to MG3 due to a smaller weighting coefficient, even when MG1 and MG2 have excessive power from the PV at daytime. Similarly, when the PV output in MG1 cannot meet all the local loads at hour 17, the higher priority has granted it to acquire additional power from MG4 than MG1 and MG2. 

With the implementation of the proposed method, the total operational cost has been not only decreased for each individual MG, but for the entire community, as shown in Table~\ref{tab:operation_cost}. Compared with direct transaction with the upstream grid, the operational cost of each MG has been reduced from 5.109\% to 21.544\%. 
On the other hand, by using the coordination strategy, the transaction cost with the upstream grid has been decreased by \$8.231 even with additional \$1.245 of transmission loss. In consequence,
the operational cost in total has been decreased by 9.474\%.
Therefore, individual MGs can benefit from the proposed EMS by reducing local operational costs, which would potentially attract external MGs for active participation into the community.

\begin{table}[!tbp]
	\centering
	\caption{Operation cost of MG community with 4 MGs}
	\label{tab:operation_cost}
	\vspace{-4pt}
	\resizebox{\columnwidth}{!}{
		\begin{tabular}{r|rrrr|r}
			\hline
			{Cost (\$)}	&MG1&	MG2&	MG3&	MG4&	total \\ \hline
			original	&5.981	&6.068	&72.233	&-10.558	&73.724 \\ \hline
			grid-side	&5.496	&5.036	&56.028	&-1.068		&65.493 \\ \hline
			community	&-0.075	&0.560	&12.513	&-11.753	&1.245 \\ \hline
			total		&5.421	&5.596	&68.542	&-12.821	&66.737 \\ \hline
			improvement	&9.361\%&7.619\%&5.109\%&21.544\%	&9.474\% \\ \hline
		\end{tabular}
	}
	\vspace{-4pt}
\end{table}

\begin{table}[!tbp]
	\centering
	\caption{Average operation cost of community with 50 MGs}
	\label{tab:operation_cost_case50}
	\vspace{-4pt}
	\resizebox{\columnwidth}{!}{
		\begin{tabular}{r|rrrr|r}
			\hline
			{Cost (\$)}	&MG1&	MG2&	MG3&	MG4&	total \\ \hline
			original	&6.083	&6.035	&71.979	&-10.487	&548.361\\ \hline
			grid-side	&5.503	&5.040	&55.770	&-1.059		&490.268 \\ \hline
			community	&0.208	&0.560	&12.306	&-10.996	&8.036 \\ \hline
			total		&5.711	&5.600	&68.076	&-12.055	&498.304 \\ \hline
			improvement	&6.111\%&7.201\%&5.422\%&14.958\%&9.129\% \\  \hline
		\end{tabular}
	}
	\vspace{-4pt}
\end{table}

\subsection{Case 2: 50-MG community}
A regional MG community with 50 MGs in four types including 20 of MG1, 20 of MG2, 5 of MG3 and 5 of MG4, respectively, is further investigated to validate the scalability of the proposed EMS.
Individual MG characteristics follow those of same types in Case 1.

The operational cost results are presented in Table~\ref{tab:operation_cost_case50}. It is observed the entire MG community has reduced the operational cost in 24 hours from \$548.361 to \$490.268, while the total transmission loss is \$8.036. In total, the electricity expense has been improved by 9.129\%. 
In addition, the proposed EMS benefits all types of individual MGs with average cost improvements ranging from 5.422\% to 14.958\%. As a result, the implementation of the proposed EMS may implicitly decrease the electricity price from the upstream grid in turn, since the stress to power congestion at peak hours in the distribution level would be alleviated. 

Comparing the result with that in the 4-MG case, it is noted that the ratio of operational result reduction does not necessarily decrease with the increasing size of MG community, since it is closely related on the different MG types and their independent operational states, such as RES outputs, fixed and dispatchable load profiles, electricity prices and so on.

\subsection{Comparison with Other Methods}
Three existing approaches in the literature are conducted for comparison to further evaluate the performance and advantages of the proposed EMS.
Firstly, the single-level energy management framework in \cite{RN59} is adopted as the centralized benchmark in which all the MGs in the community are combined just as one unity. A mathematical model similar in Section~\ref{subsec:constraints} is formulated and solved. 
Secondly, the optimization method proposed in \cite{papa2014decen} without stochastic processes is utilized in the community level as the two-level EMS benchmark, in which a Lagrangian relaxation based algorithm is used to eliminate power exchange imbalances. 
Finally, a purely decentralized framework by only using local EMSs is added in which each MG makes transaction directly with the upstream grid, without any communication inside the community. 

\begin{table}[!tbp]
	\centering
	\caption{Comparative results for 4 MGs} \label{tab:case_comparison_4}
	\vspace{-4pt}
	\resizebox{0.9\columnwidth}{!}{
		\begin{tabular}{c|c|c}
			\hline
			{Algorithm}&	Operational cost (\$)&	Computation time (s) \\\hline
			proposed&	66.73&	1.17 \\\hline
			\cite{RN59}&	66.60&	36.16 \\\hline
			\cite{papa2014decen}&	66.67&	22.49 \\ \hline
			direct transaction &	73.72&	1.01 \\
			\hline
		\end{tabular}
	}
\end{table}

\begin{table}[!tbp]
	\centering
	\caption{Comparative results for 50 MGs} \label{tab:case_comparison_50}
	\vspace{-4pt}
	\resizebox{0.9\columnwidth}{!}{
		\begin{tabular}{c|c|c}
			\hline
			{Algorithm}&	Operational cost (\$)&	Computation time (s) \\\hline
			proposed	&498.30		&15.06 \\	\hline
			\cite{RN59}	&497.22		&2053.48 \\	\hline
			\cite{papa2014decen}	&497.96	&1378.13 \\ \hline
			direct transaction 	&617.41	&12.82 \\
			\hline
		\end{tabular}
	}
\end{table}

Comparative results with existing methods in Case 1 and 2 are presented in Table~\ref{tab:case_comparison_4} and Table~\ref{tab:case_comparison_50}, respectively. 
It is seen in Table~\ref{tab:case_comparison_4} that 
the proposed EMS has achieved the optimized operational cost slightly higher by just 0.2\% to the best result with \cite{RN59}, whereas the computation time is nearly 19 times faster than the second best result with \cite{papa2014decen}. 
It is noted that the operational cost in \cite{RN59} is the minimum since all the information from individual MGs has been fully collected and gathered into the centralized EMS, which may nevertheless bring serious privacy concerns.
Similar privacy problems exist in \cite{papa2014decen} that information of dispatch signals needs to be exchanged among MGs and with the community-level EMS. On the contrary, private information is well preserved in the proposed EMS, since no communication among local EMSs or with the community-level EMS is required by any means. Communications between the community-level EMS and local EMSs only involve total power exchanges of individual MGs that do not expose any detailed decision making inside MGs. 

The benefits on computational speed of the proposed EMS are further revealed with the increasing number of MGs, as shown in Table~\ref{tab:case_comparison_50}.
It is seen that the proposed EMS spends just 15 seconds to the optimized result with a degradation of 0.21\%, however it takes nearly half an hour for the centralized method in \cite{RN59} and 23 minutes for the decentralized approach in \cite{papa2014decen}. Such the discrepancy on computational efficiency would become more noticeable with size expansion of the MG community, therefore, the proposed EMS has outstanding advantages in view of trade-offs between computation efficiency and solution quality. 
With this respect, it is noted that the uncertainty related with renewables and loads are not modeled in the proposed method because of its superior performance on computation speed. The proposed EMS is able to be carried out more frequently if uncertain factors are taken into consideration. 

\section{Conclusion} \label{sec:conclusion}
\subsection{Summary}
In this paper, a hierarchically coordinated EMS model for multiple small-scale MGs in a regional community is proposed, accounting for minimization of the community-level operational cost and maximization of individual MG-level benefits simultaneously. 
The local EMSs aim to minimize the individual operational cost, while the community-level EMS determines specific energy transactions in the community by using the pairing algorithm to further reduce individual operational costs. 
The proposed EMS has been validated by two case studies with different scales. By comparing with existing approaches, simulation results have shown significant advantages of the proposed EMS on modeling generality, computational complexity and privacy security, that 
the optimization time has been reduced significantly by the non-iterative algorithm, and privacy issues have been eliminated by minimal information exchange. 

\subsection{Future work}
There have been studies addressing uncertainties by implementing stochastic optimization techniques \cite{hu2018toward, valencia2016robust, malekpour2017stochastic}, in which uncertainties are usually handled by sampling reduction and uncertainty set in stochastic programming, and then the optimization problems can be solved in multiple stages. 
Nevertheless, most of the techniques and algorithms mentioned above have been implemented to the centralized energy management framework whereas little application has been reported regarding decentralized energy management. 
In our future research, it is planned to incorporate uncertainties into the proposed model with distributed stochastic optimization schemes, and develop computationally efficient algorithm for practical application. 

\section*{Acknowledgment}
This research is supported by the National Research Foundation, Prime Minister’s Office, Singapore under the Energy Innovation Research Programme (EIRP) Energy Storage Grant Call and administrated by the Energy Market Authority (NRF2015EWT-EIRP002-007).

\bibliographystyle{IEEEtran}
\bibliography{IEEEabrv,myConf_short,myRef}

\begin{thebibliography}{10}
\providecommand{\url}[1]{#1}
\csname url@samestyle\endcsname
\providecommand{\newblock}{\relax}
\providecommand{\bibinfo}[2]{#2}
\providecommand{\BIBentrySTDinterwordspacing}{\spaceskip=0pt\relax}
\providecommand{\BIBentryALTinterwordstretchfactor}{4}
\providecommand{\BIBentryALTinterwordspacing}{\spaceskip=\fontdimen2\font plus
\BIBentryALTinterwordstretchfactor\fontdimen3\font minus
  \fontdimen4\font\relax}
\providecommand{\BIBforeignlanguage}[2]{{%
\expandafter\ifx\csname l@#1\endcsname\relax
\typeout{** WARNING: IEEEtran.bst: No hyphenation pattern has been}%
\typeout{** loaded for the language `#1'. Using the pattern for}%
\typeout{** the default language instead.}%
\else
\language=\csname l@#1\endcsname
\fi
#2}}
\providecommand{\BIBdecl}{\relax}
\BIBdecl

\bibitem{kanchev2011energy}
H.~Kanchev, D.~Lu, F.~Colas, V.~Lazarov, and B.~Francois, ``Energy management
  and operational planning of a microgrid with a pv-based active generator for
  smart grid applications,'' \emph{IEEE transactions on industrial
  electronics}, vol.~58, no.~10, pp. 4583--4592, 2011.

\bibitem{anvari2017efficient}
A.~Anvari-Moghaddam, J.~M. Guerrero, J.~C. Vasquez, H.~Monsef, and
  A.~Rahimi-Kian, ``Efficient energy management for a grid-tied residential
  microgrid,'' \emph{IET Generation, Transmission \& Distribution}, vol.~11,
  no.~11, pp. 2752--2761, 2017.

\bibitem{liu2015heuristic}
N.~Liu, Q.~Chen, J.~Liu, X.~Lu, P.~Li, J.~Lei, and J.~Zhang, ``A heuristic
  operation strategy for commercial building microgrids containing evs and pv
  system,'' \emph{IEEE Transactions on Industrial Electronics}, vol.~62, no.~4,
  pp. 2560--2570, 2015.

\bibitem{RN397}
X.~Liu, P.~Wang, and P.~C. Loh, ``A hybrid {AC/DC} microgrid and its
  coordination control,'' \emph{{IEEE} Trans. Smart Grid}, vol.~2, no.~2, pp.
  278--286, 2011.

\bibitem{olivares2014trends}
D.~E. Olivares, A.~Mehrizi-Sani, A.~H. Etemadi, C.~A. Ca{\~n}izares,
  R.~Iravani, M.~Kazerani, A.~H. Hajimiragha, O.~Gomis-Bellmunt, M.~Saeedifard,
  R.~Palma-Behnke \emph{et~al.}, ``Trends in microgrid control,'' \emph{{IEEE}
  Trans. Smart Grid}, vol.~5, no.~4, pp. 1905--1919, 2014.

\bibitem{fathi2013sta}
M.~Fathi and H.~Bevrani, ``Statistical cooperative power dispatching in
  interconnected microgrids,'' \emph{{IEEE} Trans. Softw. Eng.}, vol.~4, no.~3,
  pp. 586--593, 2013.

\bibitem{chiu2015mul}
W.-Y. Chiu, H.~Sun, and H.~V. Poor, ``A multiobjective approach to
  multimicrogrid system design,'' \emph{{IEEE} Trans. Smart Grid}, vol.~6,
  no.~5, pp. 2263--2272, 2015.

\bibitem{zhang2017scopf}
W.~Zhang, Y.~Xu, Z.~Dong, and K.~P. Wong, ``Robust security constrained-optimal
  power flow using multiple microgrids for corrective control of power systems
  under uncertainty,'' \emph{{IEEE} Trans. Ind. Informat.}, vol.~13, no.~4, pp.
  1704--1713, 2017.

\bibitem{7463503}
S.~Chanda and A.~K. Srivastava, ``Defining and enabling resiliency of electric
  distribution systems with multiple microgrids,'' \emph{{IEEE} Trans. Smart
  Grid}, vol.~7, no.~6, pp. 2859--2868, Nov. 2016.

\bibitem{sugg_e}
B.~Zhao, X.~Wang, D.~Lin, M.~M. Calvin, J.~C. Morgan, R.~Qin, and C.~Wang,
  ``Energy management of multiple microgrids based on a system of systems
  architecture,'' \emph{{IEEE} Trans. Power Syst.}, vol.~33, no.~6, pp.
  6410--6421, 2018.

\bibitem{sugg_d}
F.~Luo, G.~Ranzi, S.~Wang, and Z.~Y. Dong, ``Hierarchical energy management
  system for home microgrids,'' \emph{{IEEE} Trans. Smart Grid}, 2018.

\bibitem{sugg_b}
F.~Luo, G.~Ranzi, C.~Wan, Z.~Xu, and Z.~Y. Dong, ``A multistage home energy
  management system with residential photovoltaic penetration,'' \emph{{IEEE}
  Trans. Ind. Informat.}, vol.~15, no.~1, pp. 116--126, 2019.

\bibitem{RN394}
N.~G. Paterakis, O.~Erdin{\c{c}}, I.~N. Pappi, A.~G. Bakirtzis, and J.~P.~S.
  Catal{\~a}o, ``Coordinated operation of a neighborhood of smart households
  comprising electric vehicles, energy storage and distributed generation,''
  \emph{{IEEE} Trans. Smart Grid}, vol.~PP, no.~99, pp. 1--12, 2016.

\bibitem{7042324}
N.~Nikmehr and S.~N. Ravadanegh, ``Optimal power dispatch of multi-microgrids
  at future smart distribution grids,'' \emph{{IEEE} Trans. Smart Grid},
  vol.~6, no.~4, pp. 1648--1657, Jul. 2015.

\bibitem{zhang2017robust}
C.~Zhang, Y.~Xu, Z.~Y. Dong, and K.~P. Wong, ``Robust coordination of
  distributed generation and price-based demand response in microgrids,''
  \emph{{IEEE} Trans. Smart Grid}, vol.~9, no.~5, pp. 4236--4247, 2018.

\bibitem{RN389}
J.~Ni and Q.~Ai, ``Economic power transaction using coalitional game strategy
  in micro-grids,'' \emph{IET Gener. Transm. Distrib.}, vol.~10, no.~1, pp.
  10--18, 2016.

\bibitem{RN392}
J.~Li, Y.~Liu, and L.~Wu, ``Optimal operation for community based multi-party
  microgrid in grid-connected and islanded modes,'' \emph{{IEEE} Trans. Smart
  Grid}, vol.~9, no.~2, pp. 756--765, 2018.

\bibitem{wang2015coordinated}
Z.~Wang, B.~Chen, J.~Wang, M.~M. Begovic, and C.~Chen, ``Coordinated energy
  management of networked microgrids in distribution systems,'' \emph{{IEEE}
  Trans. Smart Grid}, vol.~6, no.~1, pp. 45--53, 2015.

\bibitem{sugg_c}
S.~A. Arefifar, M.~Ordonez, and Y.~A.-R.~I. Mohamed, ``Energy management in
  multi-microgrid systems-{D}evelopment and assessment,'' \emph{{IEEE} Trans.
  Power Syst.}, vol.~32, no.~2, pp. 910--922, 2017.

\bibitem{ouammi2015coordinated}
A.~Ouammi, H.~Dagdougui, L.~Dessaint, and R.~Sacile, ``Coordinated model
  predictive-based power flows control in a cooperative network of smart
  microgrids,'' \emph{{IEEE} Trans. Smart Grid}, vol.~6, no.~5, pp. 2233--2244,
  2015.

\bibitem{sugg_a}
B.~Celik, R.~Roche, D.~Bouquain, and A.~Miraoui, ``Decentralized neighborhood
  energy management with coordinated smart home energy sharing,'' \emph{{IEEE}
  Trans. Smart Grid}, vol.~9, no.~6, pp. 6387--6397, 2018.

\bibitem{papa2014decen}
D.~Papadaskalopoulos, D.~Pudjianto, and G.~Strbac, ``Decentralized coordination
  of microgrids with flexible demand and energy storage,'' \emph{{IEEE} Trans.
  Sustain. Energy}, vol.~5, no.~4, pp. 1406--1414, 2014.

\bibitem{TR126}
S.~Drouilhet, B.~Johnson, S.~Drouilhet, and B.~Johnson, ``A battery life
  prediction method for hybrid power applications,'' in \emph{35th Aero. Sci.s
  Meet. and Exhib.}, 1997, p. 948.

\bibitem{ju2017two}
C.~Ju, P.~Wang, L.~Goel, and Y.~Xu, ``A two-layer energy management system for
  microgrids with hybrid energy storage considering degradation costs,''
  \emph{{IEEE} Trans. Smart Grid}, vol.~9, no.~6, pp. 6047--6057, Nov. 2018.

\bibitem{ding2017new}
T.~Ding, Y.~Lin, G.~Li, and Z.~Bie, ``A new model for resilient distribution
  systems by microgrids formation,'' \emph{{IEEE} Trans. Power Syst.}, 2017.

\bibitem{chen2016resilient}
C.~Chen, J.~Wang, F.~Qiu, and D.~Zhao, ``Resilient distribution system by
  microgrids formation after natural disasters,'' \emph{{IEEE} Trans. Smart
  Grid}, vol.~7, no.~2, pp. 958--966, 2016.

\bibitem{kashem2000novel}
M.~A. Kashem, V.~Ganapathy, G.~B. Jasmon, and M.~I. Buhari, ``A novel method
  for loss minimization in distribution networks,'' in \emph{2000 IEEE Elect.
  Util. Deregu. and Restr. and Power Techno.}, 2000, pp. 251--256.

\bibitem{jabr2012minimum}
R.~A. Jabr, R.~Singh, and B.~C. Pal, ``Minimum loss network reconfiguration
  using mixed-integer convex programming,'' \emph{{IEEE} Trans. Power Syst.},
  vol.~27, no.~2, pp. 1106--1115, 2012.

\bibitem{turitsyn2011options}
K.~Turitsyn, P.~Sulc, S.~Backhaus, and M.~Chertkov, ``Options for control of
  reactive power by distributed photovoltaic generators,'' \emph{Proceedings of
  the IEEE}, vol.~99, no.~6, pp. 1063--1073, 2011.

\bibitem{jabr2018linear}
R.~A. Jabr, ``Linear decision rules for control of reactive power by
  distributed photovoltaic generators,'' \emph{IEEE Transactions on Power
  Systems}, vol.~33, no.~2, pp. 2165--2174, 2018.

\bibitem{rgurobi}
\BIBentryALTinterwordspacing
I.~Gurobi~Optimization, ``Gurobi optimizer reference manual,'' 2016. [Online].
  Available: \url{http://www.gurobi.com}
\BIBentrySTDinterwordspacing

\bibitem{SERIS}
\BIBentryALTinterwordspacing
``Solar irradiance map,'' 2017. [Online]. Available:
  \url{https://www.solar-repository.sg/solar-irradiance-map}
\BIBentrySTDinterwordspacing

\bibitem{EMASG}
\BIBentryALTinterwordspacing
``Singapore energy statistics,'' 2017. [Online]. Available:
  \url{https://www.ema.gov.sg/singapore_energy_statistics.aspx}
\BIBentrySTDinterwordspacing

\bibitem{RN207}
\BIBentryALTinterwordspacing
``Price {I}nformation,'' 2017. [Online]. Available:
  \url{https://www.emcsg.com/marketdata}
\BIBentrySTDinterwordspacing

\bibitem{RN59}
D.~E. Olivares, C.~A. Canizare, and M.~Kazerani, ``A centralized energy
  management system for isolated microgrids,'' \emph{{IEEE} Trans. Smart Grid},
  vol.~5, no.~4, pp. 1864--1875, 2014.

\bibitem{hu2018toward}
W.~Hu, P.~Wang, and H.~B. Gooi, ``Toward optimal energy management of
  microgrids via robust two-stage optimization,'' \emph{{IEEE} Trans. Smart
  Grid}, vol.~9, no.~2, pp. 1161--1174, 2018.

\bibitem{valencia2016robust}
F.~Valencia, D.~S{\'a}ez, J.~Collado, F.~{\'A}vila, A.~Marquez, and J.~J.
  Espinosa, ``Robust energy management system based on interval fuzzy models,''
  \emph{{IEEE} Trans. Control Syst. Technol.}, vol.~24, no.~1, pp. 140--157,
  2016.

\bibitem{malekpour2017stochastic}
A.~R. Malekpour and A.~Pahwa, ``Stochastic networked microgrid energy
  management with correlated wind generators,'' \emph{{IEEE} Trans. Power
  Syst.}, vol.~32, no.~5, pp. 3681--3693, 2017.

\end{thebibliography}

\end{document}